\documentclass[journal]{IEEEtran}
\usepackage{amsmath,amsfonts}
\usepackage{algorithmic}
\usepackage{algorithm}
\usepackage{array}
\usepackage{textcomp}
\usepackage{stfloats}
\usepackage{url}
\usepackage{verbatim}
\usepackage{graphicx}
\usepackage{cite}
\usepackage{xcolor}
\usepackage{multirow}
\usepackage{makecell}
\usepackage{amsthm}
\usepackage{subfigure}
\usepackage{booktabs}

\begin{document}

\title{Disentangled Information Bottleneck guided Privacy-Protective JSCC for Image Transmission\vspace{0.3cm}}

\author{Lunan Sun, Yang Yang, \emph{Member, IEEE}, Mingzhe Chen, \emph{Member, IEEE}, Caili Guo, \emph{Senior Member, IEEE}}
\maketitle

\begin{abstract}
Joint source and channel coding (JSCC) has attracted increasing attention due to its robustness and high efficiency. However, JSCC is vulnerable to privacy leakage due to the high relevance between the source image and channel input. In this paper, we propose a disentangled information bottleneck guided privacy-protective JSCC (DIB-PPJSCC) for image transmission, which aims at protecting private information and achieving superior communication performance. In particular, we propose a DIB objective to compress the private information in public subcodewords and improve the reconstruction quality simultaneously. To optimize JSCC neural networks using the DIB objective, we derive a differentiable estimation based on variational approximation and the density-ratio trick. Additionally, we design a password-based privacy-protective algorithm that encrypts the private subcodewords, achieving joint optimization with JSCC neural networks. The proposed algorithm involves an encryptor for encrypting private information and a decryptor for recovering it at the legitimate receiver. A loss function is derived based on the maximum entropy principle for jointly training the encryptor, decryptor, and JSCC decoder to maximize eavesdropping uncertainty and improve reconstruction quality. Experimental results show that DIB-PPJSCC reduces eavesdropping accuracy on private information by up to 15\% and decreases inference time by 10\%.

\end{abstract}


\section{Introduction}\label{Introduction}
Shannon's information theory has laid the cornerstone of modern communication systems. In particular, Shannon’s separation theorem pointed out that separate source and channel coding (SSCC) is optimal for a memoryless source and channel when the latency, complexity, and code length are not constrained\cite{shannon1948mathematical}. However, the assumption of potentially infinite code lengths is impractical in real-world scenarios, and thus SSCC is suboptimal for finite code lengths. Additionally, to achieve theoretically optimal performance, maximum likelihood detection methods must be used, which is generally NP-hard\cite{berlekamp1978inherent}, thus introducing high computational complexity and leading to unacceptable latency. Moreover, the envisioned sixth generation (6G) of wireless networks are expected to support a wide range of services and applications, such as ugmented reality, medical imaging and autonomous vehicles\cite{chafii2022ten}, which have strict latency requirement\cite{chen2021distributed}. Therefore, SSCC may not be able to meet the demands of 6G. 

To address these challenges, joint source and channel coding (JSCC) has attracted increasing attention as a means to achieve reliable data transmission. In the initial stage, the studies on JSCC focus on coding schemes \cite{cheung2000bit, heinen2005transactions} or performance analysis under ideal channel or source assumptions\cite{gallager1968information,gastpar2003code}. However, these hand-crafted coding schemes may require additional tuning. Recently, deep learning (DL) based approaches have been proposed for JSCC, driven by the rapid advancements in artificial intelligence (AI). Specifically, since images possess larger dimensions compared to speech and text data, there exists more information redundancy in images, and transmitting image data requires a higher rate than transmitting speech and text data. Therefore, DL-based JSCC system for image transmissions has attracted a plethora of prior art, and these works have exhibited appealing properties in terms of image restoration quality\cite{bourtsoulatze2019deep, choi2019neural, sun2023adaptive}, flexibility\cite{xu2021wireless}, and robustness\cite{song2020infomax}. 

However, the aforementioned studies mainly aim at improving the system performance, while ignoring the potential adversarial eavesdroppers and privacy leakage during image transmission. To enhance security, recent studies combined existing hand-crafted encryption schemes with DL-based JSCC to encrypt the transmitted information\cite{xu2021deep,tung2022deep}. However, the introduction of encryption schemes may lead to high computational complexity and a reduction of image reconstruction quality\cite{massoudi2008overview}. Some other studies on JSCC for image transmission have taken into account the protection of specific private information\cite{Marchioro2020Adversarial, Erdemir2022Privacy} by adversarially training the JSCC encoder/decoder and the hypothetical eavesdropper to discard the private information in the codeword before transmission. However, the loss of private information can lead to significant degradation in terms of the image reconstruction quality. Therefore, a privacy-protective JSCC that not only safeguards private information but also achieves superior communication performance at the legitimate receiver deserves investigation.

\subsection{Related work and Motivations}
The existing works on DL-based JSCC for image transmission typically model the communication system as an end-to-end deep neural network (DNN)-based autoencoder (AE)\cite{bourtsoulatze2019deep,sun2023adaptive,kurka2020deepjscc,xu2021wireless,choi2019neural,song2020infomax,xu2021deep,tung2022deep, Marchioro2020Adversarial, Erdemir2022Privacy}. The seminal work \cite{bourtsoulatze2019deep} proposed an AE-based JSCC architecture called deep JSCC to minimize the reconstruction distortion, which outperforms conventional SSCC under the Gaussian channel and slow Rayleigh fading channel. Based on \cite{bourtsoulatze2019deep}, the authors in \cite{kurka2020deepjscc} exploited the channel feedback and further improved the reconstruction quality of deep JSCC. To deal with the performance degradation caused by the variations of signal-to-noise ratios (SNRs), the authors in \cite{xu2021wireless} employed a channel-wise soft attention network to adapt to varying channel conditions. The above studies concentrate on DL-based JSCC with the Gaussian channel and the Rayleigh channel. For the discrete channel such as the binary symmetric channel (BSC), the authors in \cite{choi2019neural} designed a discrete variational autoencoder (VAE)-based deep JSCC model and maximized the mutual information between the source and noisy codewords. The authors in \cite{song2020infomax} further developed a JSCC model to maximize the mutual information between the source and the codewords based on \cite{choi2019neural}. Both works in \cite{choi2019neural} and \cite{song2020infomax} focused on minimizing the reconstruction distortion while ignoring the required rate. In contrast, the authors in \cite{sun2023adaptive} proposed an adaptive information bottleneck (IB) guided JSCC which jointly minimized the distortion and the rate, aligning with the rate-distortion theory. 

Even though interesting, these DL-based JSCC studies for image transmission aim at reducing reconstruction errors while ignoring the security and privacy aspects of JSCC. Recently, the authors in \cite{xu2021deep} considered the joint encryption and source-channel coding and placed the encryption module in front of the JSCC encoder. The raw images were directly encrypted using the discrete cosine transform (DCT) and then input into the JSCC encoder. The authors in \cite{tung2022deep} followed a different approach by positioning the encryption process behind the JSCC encoder and employing a key cryptographic scheme to encrypt the output of the JSCC encoder. In \cite{xu2021deep}, all pixels in the raw image were encrypted, while in \cite{tung2022deep}, all the extracted semantic information from the raw image was encrypted. Both methods encrypted the entire information of the raw image. In fact, the privacy protection goal is to protect specific private information rather than entire information mostly, and encrypting entire information can lead to extra computation and storage costs\cite{Roy_2019_CVPR, BOULEMTAFES202021}. Moreover, encrypting and decrypting entire information may result in a larger reconstruction error\cite{bloch2021overview}. To tackle this issue, the work in \cite{Marchioro2020Adversarial} first considered the DL-based JSCC that aimed at protecting specific private information and proposed an adversarially trained JSCC with Gaussian channels. The main objective of their neural networks is minimizing the reconstruction distortion while minimizing the mutual information between the private information and the noisy codewords received by the eavesdropper. Furthermore, the authors in \cite{Erdemir2022Privacy} introduced a VAE-based JSCC model for privacy protection with BSC. They jointly minimized the reconstruction distortion and the mutual information between the private information and the noisy codewords received by the eavesdropper, as well as maximizing the mutual information between the raw image and the codewords extracted by the JSCC encoder. However, these works discarded the information related to private information during image transmission. The legitimate receiver had no access to private information, thus severely damaging the reconstruction quality. 

To solve these challenges, we need to separate private information and public information and protect private information without altering public information. However, existing JSCC studies cannot distinguish between private and public information due to the inexplicability of neural networks. Recently, the authors in \cite{tishby2000information} introduced an information-theoretic principle for neural networks, termed IB, which extracts information relevant to target prediction while compressing information irrelevant to target prediction. Since IB gives an information-theoretic explanation for neural networks, it enables the neural networks to disentangle the target-relevant and the target-irrelevant information\cite{pan2021disentangled}. Therefore, we propose a novel disentangled IB guided privacy-protective JSCC (DIB-PPJSCC) that is able to disentangle private and public information. However, the information-theoretic principle in \cite{tishby2000information} is designed for image classification and does not consider communication over channel. Thus, it is not suitable for the considered scenario where images are transmitted over channel and recovered by the receiver. In addition, traditional encryption schemes are not suitable for protecting the private information extracted by the JSCC encoder. These schemes are designed for discrete inputs and are not easily integrated with JSCC neural networks, which generate continuous outputs. Thus, the encryption schemes hinder gradient backpropagation and fail to achieve the ideal optimal solution. Therefore, we need to design a new form of disentangled IB objective as well as a privacy-protective algorithm that can be jointly optimized with the JSCC neural networks so as to achieve a balance between privacy protection and reconstruction quality.

\subsection{Contributions}
The main contribution of this paper is a DIB-PPJSCC scheme for image transmission. The major contributions of the paper can be summarized as follows:

\begin{itemize}
\item {We design a new form of DIB objective for privacy-protective JSCC that aims at maximizing the mutual information between the private subcodewords and the private information, minimizing the reconstruction distortion as well as minimizing the mutual information between the private and the public subcodewords. Thus, the new DIB objective enables privacy-protective JSCC to disentangle private and public information into private and public subcodewords which are independent of each other. To the best of the authors' knowledge, this is the first work that introduces the IB principle to design privacy-protective JSCC for image transmission.}

\item{The mutual information used in the DIB objective is intractable for DNNs with high-dimensional representations. Therefore, we develop a new mathematically tractable and differentiable estimation of mutual information via variational approximations and density ratio trick by involving a discriminator and a classifier to estimate posterior probability. These derived estimations are used as the loss function of DIB-PPJSCC.}

\item {We propose a privacy-protective algorithm, which can be optimized with the JSCC neural networks jointly in order to protect the private subcodewords during transmission and recover private information at the legitimate receiver. Specifically, we apply a private information encryptor and decryptor to achieve a password-based protection process. To jointly optimize the encryptor, decryptor and JSCC decoder, we derive a novel loss function based on the maximum entropy principle, which aims at maximizing the randomness of the eavesdropping as well as improving the reconstruction quality.}

\end{itemize}

We compare DIB-PPJSCC with state-of-the-art privacy-protective JSCC methods and the traditional separate scheme. Simulation results show that compared with baselines, DIB-PPJSCC significantly reduces the reconstruction error, the eavesdropping accuracy on private information as well as the complexity. 

The rest of this paper is organized as follows. In Section II, the system model is described. The DIB objective for privacy-protective JSCC and its optimization are presented in Section III. The privacy-protective algorithm is introduced in Section IV. Extensive experimental results to verify the effectiveness of DIB-PPJSCC are provided in Section V. Finally, the conclusions are drawn in Section VI.

\section{System Model}\label{system-model}
\begin{figure}[t]
\centering{
\includegraphics[width=1\columnwidth]{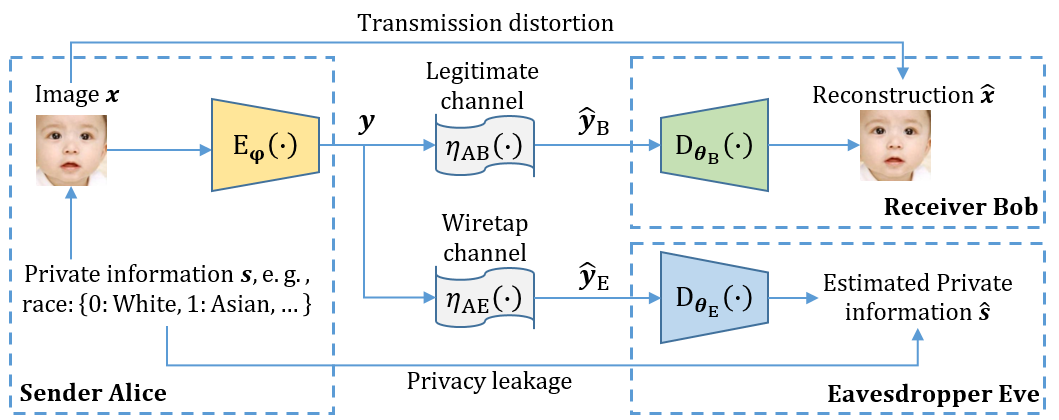}
}
\caption{\label{fig:illustration}An illustration of the privacy-protective JSCC system with wiretap channel.}
\end{figure}

As shown in Fig. \ref{fig:illustration}, we consider a communication scenario where a sender Alice transmits an image $\boldsymbol{x} \in {\mathbb{R}^{N}}$ with size $N = H$ (Height) $\times$ $W$ (Width) $\times$ $C$ (Channel) to a legitimate receiver Bob, where $\mathbb{R}$ represents the set of real numbers. The image $\boldsymbol{x}$ contains a certain class of private information, which is encoded into one-hot vector ${\boldsymbol{s}} \in {\left\{ {0,1} \right\}^S}$, where $S$ represents the number of classes of the private information. Alice encodes the image $\boldsymbol{x}$ into a codeword $\boldsymbol{y} \in {\mathbb{R}^{M}}$, where $M$ represents the length of the codeword $\boldsymbol{y}$ to be transmitted. The encoding function ${{\rm{E}}_{\boldsymbol{\varphi }}}:{{\mathbb{R}^{N}}} \to {{\mathbb{R}^{M}}}$ is parameterized by an encoder neural network with parameters $\boldsymbol{\varphi } $, and the encoding process can be expressed as:
\begin{equation}
\label{model-0}
\boldsymbol{y} = {{\rm{E}}_{\boldsymbol{\varphi }} }\left( \boldsymbol{x} \right),
\end{equation}
The codeword $\boldsymbol{y}$ is transmitted across a noisy channel ${\eta _{{\rm{AB}}}}:{\mathbb{R}^{M}} \to {\mathbb{R}^{M}}$. We consider the widely used Additive White Gaussian Noise (AWGN) wiretap channel model. The channel output noisy codeword $\boldsymbol{\hat y}_{\rm{B}} \in {\mathbb{R}^{M}}$ received by Bob is
\begin{equation}
\label{model-1}
{{{\boldsymbol{\hat y}}}_{\rm{B}}} = {\eta _{{\rm{AB}}}}\left( {\boldsymbol{y}} \right) = {\boldsymbol{y}} + {{\boldsymbol{z}}_{\rm{B}}},
\end{equation}
where ${{\boldsymbol{z}}_{\rm{B}}} \sim \mathcal{N}\left( {0,\sigma _{\rm{B}}^2{\boldsymbol{I}}} \right)$ represents the additive white Gaussian noise at Bob. At the same time, an external eavesdropper Eve has access to the codeword $\boldsymbol{y}$ via an eavesdropping channel ${\eta _{{\rm{AE}}}}:{\mathbb{R}^{M}} \to {\mathbb{R}^{M}}$. The channel output noisy codeword $\boldsymbol{\hat y}_{\rm{E}} \in {\mathbb{R}^{M}}$ received by Eve is
\begin{equation}
\label{model-2}
{{{\boldsymbol{\hat y}}}_{\rm{E}}} = {\eta _{{\rm{AE}}}}\left( {\boldsymbol{y}} \right) = {\boldsymbol{y}} + {{\boldsymbol{z}}_{\rm{E}}},
\end{equation}
where ${{\boldsymbol{z}}_{\rm{E}}} \sim \mathcal{N}\left( {0,\sigma _{\rm{E}}^2{\boldsymbol{I}}} \right)$ represents the additive white Gaussian noise at Eve.

Bob decodes the noisy codeword ${\boldsymbol{\hat y }}_{\rm{B}}$ into reconstructed image $\boldsymbol{\hat x } \in {\mathbb{R}^{N}}$. The decoding function is parameterized by the decoder neural network parameters $\boldsymbol{\theta}_{\rm{B}}$, and the decoding process is expressed as ${{\rm{D}}_{\boldsymbol{\theta}_{\rm{B}}} }:{\mathbb{R}^{M}} \to {\mathbb{R}^{N}}$. The reconstructed image $\boldsymbol{\hat x }$ is
\begin{equation}
\label{model-3}
{\boldsymbol{\hat x}} = {{\rm{D}}_{{{\boldsymbol{\theta }}_{\rm{B}}}}}\left( {{{{\boldsymbol{\hat y}}}_{\rm{B}}}} \right) = {{\rm{D}}_{\boldsymbol{\theta }_{\rm{B}}}}\left( {{\eta _{{\rm{AB}}}}\left( {{{\rm{E}}_{\boldsymbol{\varphi }}}\left( {\boldsymbol{x}} \right)} \right)} \right).
\end{equation}

Meanwhile, Eve estimates the private information $\boldsymbol {s}$ contained in $\boldsymbol {x}$ from the received codeword ${\boldsymbol{\hat y }}_{\rm{E}}$ using its own neural network with parameter ${\boldsymbol{\theta_{\rm{E}}}}$, ${{\rm{D}}_{\boldsymbol{\theta_{\rm{E}}}} }:{\mathbb{R}^{M}} \to {\left\{ {0,1} \right\}^S}$. The estimated private information ${\boldsymbol{\hat s}} \in {\left\{ {0,1} \right\}^S}$ at Eve is
\begin{equation}
\label{model-4}
{\boldsymbol{\hat s}} = {{\rm{D}}_{{{\boldsymbol{\theta }}_{\rm{E}}}}}\left( {{{{\boldsymbol{\hat y}}}_{\rm{E}}}} \right) = {{\rm{D}}_{{{\boldsymbol{\theta }}_{\rm{E}}}}}\left( {{\eta _{{\rm{AE}}}}\left( {{{\rm{E}}_{\boldsymbol{\varphi }}}\left( {\boldsymbol{x}} \right)} \right)} \right).
\end{equation}

The goal of the considered system is to determine the encoder and decoder parameters ${\boldsymbol{\varphi }}$ and ${{\boldsymbol{\theta }}_{\rm{B}}}$ that minimize the average reconstruction error between $\boldsymbol{x}$ and $\boldsymbol{\hat x }$ (distortion) while keeping the estimated private information ${{\boldsymbol{\hat s}}}$ at Eve different from the private information $\boldsymbol{s}$ in raw image (privacy).

\begin{figure*}[t!]
\centering{
\includegraphics[width=14cm]{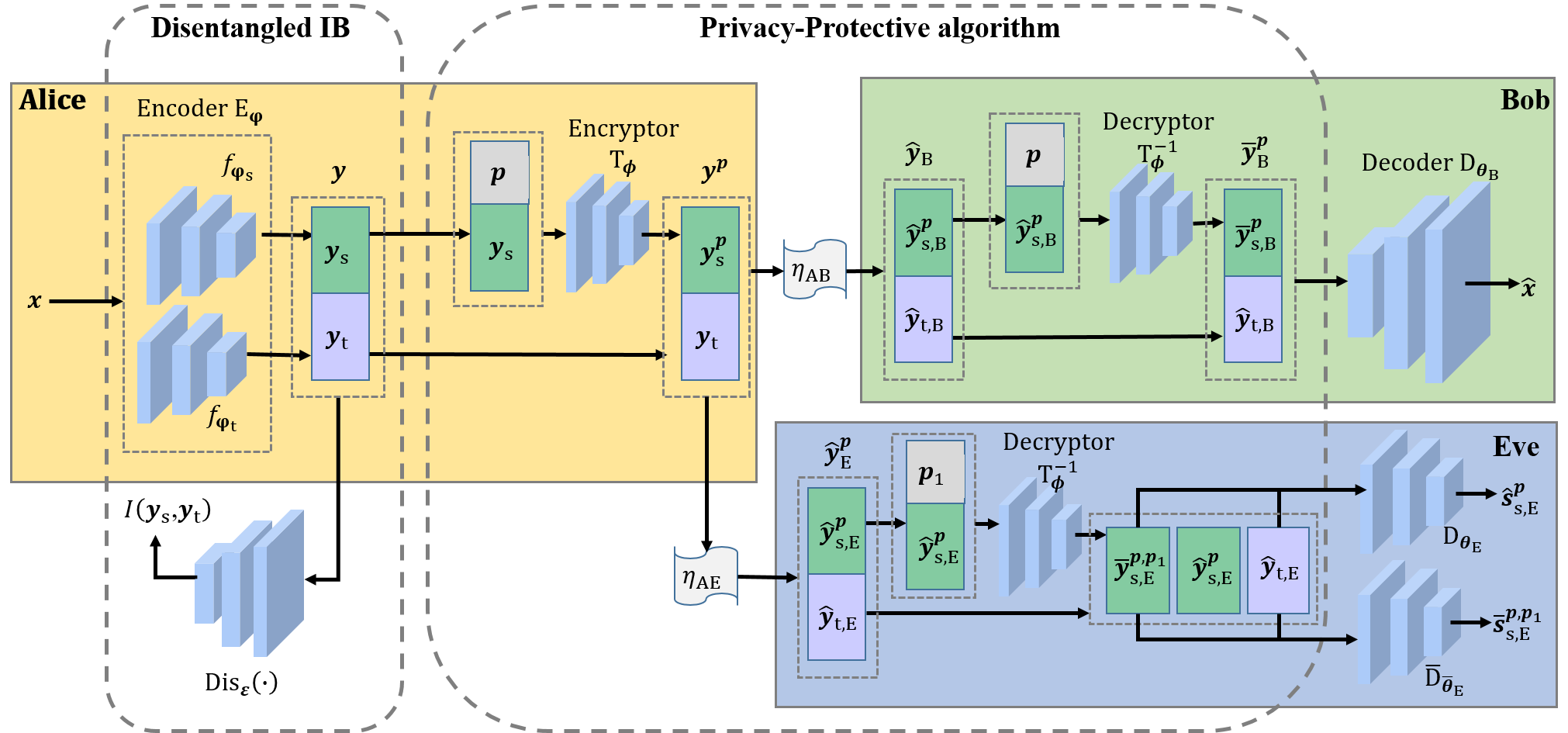}
}
\caption{\label{fig:DIB-PPJSCC}An illustration of our proposed DIB-PPJSCC. Orange: the disentangled encoder ${{\rm{E}}_{\bf{\varphi }}}$ and the encryptor ${{\rm{T}}_{\bf{\phi}} }$ at Alice. Green: the decryptor ${\rm{T}}_{\bf{\phi}} ^{ - 1}$ and the decoder ${{\rm{D}}_{{{\bf{\theta }}_{\rm{B}}}}}$ at Bob. Blue: the classifiers ${{\rm{D}}_{{{\bf{\theta }}_{\rm{E}}}}}$ and ${{{\rm{\bar D}}}_{{{{\bf{\bar \theta }}}_{\rm{E}}}}}$ at Eve. First, we train ${{\rm{E}}_{\bf{\varphi }}}$ by optimizing the DIB objective. Then, we fix ${{\rm{E}}_{\bf{\varphi }}}$ and train ${{\rm{T}}_{\bf{\phi}} }$, ${\rm{T}}_{\bf{\phi}} ^{ - 1}$ and ${{\rm{D}}_{{{\bf{\theta }}_{\rm{B}}}}}$ to defense against the eavesdropping of ${{\rm{D}}_{{{\bf{\theta }}_{\rm{E}}}}}$ and ${{{\rm{\bar D}}}_{{{{\bf{\bar \theta }}}_{\rm{E}}}}}$ according to the PP algorithm. }
\end{figure*}

\section{Distengled IB guided JSCC}\label{DIB-JSCC}
This section first introduces the proposed DIB objective for the image transmission privacy-protective JSCC system. To obtain a tractable and differentiable form of DIB objective, we then derive the estimations on mutual information terms used in DIB objective.

\subsection{DIB objective}
The considered JSCC system is shown in Fig. \ref{fig:DIB-PPJSCC}, and it mainly consists of an encoder block ${{\rm{E}}_{\boldsymbol{\varphi}} }\left(  \cdot  \right)$ at Alice, a decoder block ${{\rm{D}}_{\boldsymbol{{\theta}_{\rm{B}}}} }\left(  \cdot  \right)$ at Bob, two decoder blocks ${{\rm{D}}_{\boldsymbol{{\theta}_{\rm{E}}}} }\left(  \cdot  \right)$ and ${{{\rm{\bar D}}}_{{{{\bf{\bar \theta }}}_{\rm{E}}}}}$ at Eve, two channel blocks ${\eta _{{\rm{AB}}}}\left(  \cdot  \right)$ and ${\eta _{{\rm{AE}}}}\left(  \cdot  \right)$, and a privacy-protective algorithm block between Alice and Bob. 

In the considered system, DIB principle is adopted to guide JSCC to disentangle the codeword $\boldsymbol{y }$ into the public subcodeword and the private subcodeword which are independent with each other. Therefore, we divide the encoder block ${{\rm{E}}_{\boldsymbol{\varphi}} }\left(  \cdot  \right)$ into two components, the public encoder with parameters ${{{\boldsymbol{\varphi }}_{\rm{t}}}}$, $f_{{{\boldsymbol{\varphi }}_{\rm{t}}}}:{{\mathbb{R}}^{N}} \to {{\mathbb{R}}^{M_{\rm{t}}}}$ and the private encoder with parameters ${{{\boldsymbol{\varphi }}_{\rm{s}}}}$, ${f_{{{\boldsymbol{\varphi }}_{\rm{s}}}}}:{{\mathbb{R}}^{N}} \to {{\mathbb{R}}^{M_{\rm{s}}}}$, where ${M_{\rm{t}}}$ and ${M_{\rm{s}}}$ are the length of public subcodewords and private subcodewords, respectively. Denote the public subcodeword as  ${{\boldsymbol{y}}_{\rm{t}}} \in {\mathbb{R}^{{M_{\rm{t}}}}}$ and the private subcodewords as ${{\boldsymbol{y}}_{\rm{s}}} \in {\mathbb{R}^{{M_{\rm{s}}}}}$. We have $M = {M_{\rm{t}}} + {M_{\rm{s}}}$ and ${\boldsymbol{y}} = {\rm{concat}}\left[ {{{\boldsymbol{y}}_{\rm{t}}},{{\boldsymbol{y}}_{\rm{s}}}} \right]$, where ${\rm{concat}}\left[  \cdot  \right]$ represents the concatenation along the last axis. Then, the proposed DIB objective for separating ${{\boldsymbol{y}}_{\rm{t}}}$ and ${{\boldsymbol{y}}_{\rm{s}}}$ as well as minimize the reconstruction error is
\begin{equation}
\label{DIB-JSCC-1}
\mathop {{\rm{min}}}\limits_{{{\boldsymbol{\varphi }}_{\rm{s}}},{{\boldsymbol{\varphi }}_{\rm{t}}},{{\boldsymbol{\theta }}_{\rm{B}}}} {{\rm{E}}_{p\left( {{\boldsymbol{x,s}}} \right)}}\left( {d\left( {{\boldsymbol{x}};{\boldsymbol{\hat x}}} \right)} \right) + \alpha I\left( {{{\boldsymbol{y}}_{\rm{t}}};{{\boldsymbol{y}}_{\rm{s}}}} \right) - \beta I\left( {{{\boldsymbol{y}}_{\rm{s}}};{\boldsymbol{s}}} \right),
\end{equation}
where $\alpha $ and $\beta$ are hyperparameters, $d\left(  \cdot  \right)$ is the mean squared error (MSE) distortion of image transmission, $I\left( {{{\boldsymbol{y}}_{\rm{t}}};{{\boldsymbol{y}}_{\rm{s}}}} \right)$ is the mutual information between ${{{\boldsymbol{y}}_{\rm{t}}}}$ and ${{{\boldsymbol{y}}_{\rm{s}}}}$, and $I\left( {{{\boldsymbol{y}}_{\rm{s}}};{\boldsymbol{s}}} \right)$ is the mutual information between ${{{\boldsymbol{y}}_{\rm{s}}}}$ and ${\boldsymbol{s}}$. The first term in (\ref{DIB-JSCC-1}) is used to minimize the image transmission error when separating ${{{\boldsymbol{y}}_{\rm{t}}}}$ and ${{{\boldsymbol{y}}_{\rm{s}}}}$. The second term is used to compress the information between ${{\boldsymbol{y}}_{\rm{s}}}$ and ${{\boldsymbol{y}}_{\rm{t}}}$ thus encouraging the independence between ${{\boldsymbol{y}}_{\rm{s}}}$ and ${{\boldsymbol{y}}_{\rm{t}}}$. The third term is used to preserve the information related to $\boldsymbol{s}$ in ${{\boldsymbol{y}}_{\rm{s}}}$. By jointly minimizing $I\left( {{{\boldsymbol{y}}_{\rm{t}}};{{\boldsymbol{y}}_{\rm{s}}}} \right)$ and maximizing $I\left( {{{\boldsymbol{y}}_{\rm{s}}};{\boldsymbol{s}}} \right)$, the private information in ${{{\boldsymbol{y}}_{\rm{t}}}}$ is removed, and the public information is stored only in ${{{\boldsymbol{y}}_{\rm{t}}}}$. This ensures that ${{\boldsymbol{y}}_{\rm{t}}}$ that contains no private information, can be directly transmitted through the channel. The subsequent privacy-protective algorithm only needs to protect ${{\boldsymbol{y}}_{\rm{s}}}$ instead of the entire information. Since (\ref{DIB-JSCC-1}) compresses the private information in ${{{\boldsymbol{y}}_{\rm{t}}}}$ as well as reducing the reconstruction distortion, we refer (\ref{DIB-JSCC-1}) as DIB objective. 

However, (\ref{DIB-JSCC-1}) still cannot be applied to the privacy-protective JSCC systems, since the mutual information terms $I\left( {{{\boldsymbol{y}}_{\rm{t}}};{{\boldsymbol{y}}_{\rm{s}}}} \right)$ and $I\left( {{{\boldsymbol{y}}_{\rm{s}}};{\boldsymbol{s}}} \right)$ in (\ref{DIB-JSCC-1}) are mathematically intractable due to the unknown $p\left( {{{\boldsymbol{y}}_{\rm{t}}},{{\boldsymbol{y}}_{\rm{s}}}} \right)$, $p\left( {{{\boldsymbol{y}}_{\rm{s}}}} \right)$ and $p\left( {{{\boldsymbol{y}}_{\rm{t}}}} \right)$. To circumvent this challenge, we next derive the variational lower bound on $I\left( {{{\boldsymbol{y}}_{\rm{s}}};{\boldsymbol{s}}} \right)$ and estimation of $I\left( {{{\boldsymbol{y}}_{\rm{t}}};{{\boldsymbol{y}}_{\rm{s}}}} \right)$.

\subsection{Variational lower bound on $I\left( {{{\boldsymbol{y}}_{\rm{s}}};\boldsymbol{s}} \right)$}
Instead of maximizing the intractable true value of $I\left( {{{\boldsymbol{y}}_{\rm{s}}};{\boldsymbol{s}}} \right)$, we maximize its lower bound since the marginal and joint distributions of ${{{\boldsymbol{y}}_{\rm{s}}}}$ and ${\boldsymbol{s}}$ are unknown. According to the definition of mutual information and entropy, we have\cite{alemi2016deep}
\begin{equation}
\label{DIB-JSCC-B-1}
\begin{aligned}
I\left( {{{\boldsymbol{y}}_{\rm{s}}};{\boldsymbol{s}}} \right) & = H\left( {\boldsymbol{s}} \right) - H\left( {{\boldsymbol{s}}|{{\boldsymbol{y}}_{\rm{s}}}} \right) \ge  - H\left( {{\boldsymbol{s}}|{{\boldsymbol{y}}_{\rm{s}}}} \right)\\
& = {\mathbb{E}_{p\left( {{{\boldsymbol{y}}_{\rm{s}}},{\boldsymbol{s}}} \right)}}\left( {\log q\left( {{\boldsymbol{s}}|{{\boldsymbol{y}}_{\rm{s}}}} \right)} \right) + \underbrace {{\mathbb{E}_{p\left( {{{\boldsymbol{y}}_{\rm{s}}},{\boldsymbol{s}}} \right)}}\left( {\log \frac{{p\left( {{\boldsymbol{s}}|{{\boldsymbol{y}}_{\rm{s}}}} \right)}}{{q\left( {{\boldsymbol{s}}|{{\boldsymbol{y}}_{\rm{s}}}} \right)}}} \right)}_{{D_{{\rm{KL}}}}\left[ {p\left( {{\boldsymbol{s}}|{{\boldsymbol{y}}_{\rm{s}}}} \right)||q\left( {{\boldsymbol{s}}|{{\boldsymbol{y}}_{\rm{s}}}} \right)|} \right]  \ge 0} \\
& \ge {\mathbb{E}_{p\left( {{{\boldsymbol{y}}_{\rm{s}}},{\boldsymbol{s}}} \right)}}\left( {\log q\left( {{\boldsymbol{s}}|{{\boldsymbol{y}}_{\rm{s}}}} \right)} \right),
\end{aligned} 
\end{equation}
where $H({\boldsymbol{s}})$ is the entropy of $\boldsymbol{s}$, $H({\boldsymbol{s}}|{{{{\boldsymbol{y}}_{\rm{s}}}}})$ is the conditional entropy of $\boldsymbol{s}$ given ${{{\boldsymbol{y}}_{\rm{s}}}}$, ${q\left( {\boldsymbol{s}|{{\boldsymbol{y}}_{\rm{s}}}} \right)}$ is the variational approximation of the true posterior ${p\left( {\boldsymbol{s}|{{\boldsymbol{y}}_{\rm{s}}}} \right)}$. The equivalence in the first row of (\ref{DIB-JSCC-B-1}) is in the sense of optimization, ignoring the constant term $H\left( {\boldsymbol{s}} \right)$. ${{E_{p\left( {{{\boldsymbol{y}}_{\rm{s}}},{\boldsymbol{s}}} \right)}}\left( {\log \frac{{p\left( {{\boldsymbol{s}}|{{\boldsymbol{y}}_{\rm{s}}}} \right)}}{{q\left( {{\boldsymbol{s}}|{{\boldsymbol{y}}_{\rm{s}}}} \right)}}} \right)}$ in the second row of (\ref{DIB-JSCC-B-1}) is the KL divergence between ${q\left( {\boldsymbol{s}|{{\boldsymbol{y}}_{\rm{s}}}} \right)}$ and ${p\left( {\boldsymbol{s}|{{\boldsymbol{y}}_{\rm{s}}}} \right)}$ and is larger than $0$. Therefore, we have the third row of (\ref{DIB-JSCC-B-1}). We use a classifier ${{\rm{C}}_{\boldsymbol{\gamma }}}:{\mathbb{R}^{M}} \to {\left\{ {0,1} \right\}^S}$ consists of neural network with parameters $\boldsymbol{\gamma }$ to express ${q\left( {\boldsymbol{s}|{{\boldsymbol{y}}_{\rm{s}}}} \right)}$. To make ${q\left( {{\boldsymbol{s}}|{{\boldsymbol{y}}_{\rm{s}}}} \right)}$ close to ${p\left( {{\boldsymbol{s}}|{{\boldsymbol{y}}_{\rm{s}}}} \right)}$, we optimize ${\boldsymbol{\gamma }}$ to minimize the cross entropy between ${{\rm{C}}_{\boldsymbol{\gamma }}}\left( {{{\boldsymbol{y}}_{\rm{s}}}} \right)$ and $p\left( {{\boldsymbol{s}}|{{\boldsymbol{y}}_{\rm{s}}}} \right)$ so as to reduce the approximation error. The trained ${{\rm{C}}_{\boldsymbol{\gamma }}}\left( {{{\boldsymbol{y}}_{\rm{s}}}} \right)$ is denoted as ${q\left( {{\boldsymbol{s}}|{{\boldsymbol{y}}_{\rm{s}}}} \right)}$. Since there exists a Markov chain ${\boldsymbol{s}} \leftrightarrow {\boldsymbol{x}} \leftrightarrow {{\boldsymbol{y}}_{\rm{s}}}$, $p\left( {{{\boldsymbol{y}}_{\rm{s}}},{\boldsymbol{s}}} \right) = p\left( {{\boldsymbol{x,s}}} \right)p\left( {{{\boldsymbol{y}}_{\rm{s}}}|{\boldsymbol{x}}} \right)$. The lower bound of $I\left( {{{\boldsymbol{y}}_{\rm{s}}};{\boldsymbol{s}}} \right)$ is estimated as
\begin{equation}
\label{DIB-JSCC-B-2}
I\left( {{{\boldsymbol{y}}_{\rm{s}}};{\boldsymbol{s}}} \right) \ge {\mathbb{E}_{p\left( {{\boldsymbol{x,s}}} \right)}}{\mathbb{E}_{p\left( {{{\boldsymbol{y}}_{\rm{s}}}|{\boldsymbol{x}}} \right)}}\left[ {\log {{\rm{C}}_{\boldsymbol{\gamma }}}\left( {{{\boldsymbol{y}}_{\rm{s}}}} \right)} \right]. 
\end{equation}
As we use a deterministic ${f_{{{\boldsymbol{\varphi }}_{\rm{s}}}}}$, $p\left( {{{\boldsymbol{y}}_{\rm{s}}}|{\boldsymbol{x}}} \right)$ can be regarded as a Dirac-delta function, i.e., 
\begin{equation}
\label{DIB-JSCC-B-3}
p\left( {{{\boldsymbol{y}}_{\rm{s}}}|{\boldsymbol{x}}} \right) =
\begin{cases}
\begin{aligned}
1 & \quad \text{if } {{\boldsymbol{y}}_{\rm{s}}} = {f_{{{\boldsymbol{\varphi }}_{\rm{s}}}}}\left( {\boldsymbol{x}} \right) \\
0 & \quad \text{else}
\end{aligned}
\end{cases}.
\end{equation}
Replacing $p\left( {{{\boldsymbol{y}}_{\rm{s}}}|{\boldsymbol{x}}} \right)$ in (\ref{DIB-JSCC-B-2})
with (\ref{DIB-JSCC-B-3}), we can calculate the variational lower bound on $I\left( {{{\boldsymbol{y}}_{\rm{s}}};{\boldsymbol{s}}} \right)$ as
\begin{equation}
\label{DIB-JSCC-B-4}
I\left( {{{\boldsymbol{y}}_{\rm{s}}};{\boldsymbol{s}}} \right) \ge {\mathbb{E}_{p\left( {{\boldsymbol{x}},{\boldsymbol{s}}} \right)}}\left( {\log {{\rm{C}}_{\boldsymbol{\gamma }}}\left( {{f_{{{\boldsymbol{\varphi }}_{\rm{s}}}}}\left( {\boldsymbol{x}} \right)} \right)} \right).
\end{equation}
Then, we can use the variational lower bound in (\ref{DIB-JSCC-B-4}) as the loss function to optimize ${{\boldsymbol{\varphi }}_{\rm{s}}}$.

\subsection{Estimation of $I\left( {{{\boldsymbol{y}}_{\rm{t}}};{{\boldsymbol{y}}_{\rm{s}}}} \right)$} 
By maximizing the variational lower bound on $I\left( {{{\boldsymbol{y}}_{\rm{s}}};{\boldsymbol{s}}} \right)$, the private information can be converged in ${{\boldsymbol{y}}_{\rm{s}}}$. It is also crucial to minimize $I\left( {{{\boldsymbol{y}}_{\rm{t}}};{{\boldsymbol{y}}_{\rm{s}}}} \right)$ to enforce independence between ${{\boldsymbol{y}}_{\rm{s}}}$ and ${{\boldsymbol{y}}_{\rm{t}}}$ and prevent any private information from leaking into ${{\boldsymbol{y}}_{\rm{t}}}$. However, minimizing $I\left( {{{\boldsymbol{y}}_{\rm{t}}};{{\boldsymbol{y}}_{\rm{s}}}} \right) = {D_{{\rm{KL}}}}\left[ {p\left( {{{\boldsymbol{y}}_{\rm{t}}},{{\boldsymbol{y}}_{\rm{s}}}} \right)||p\left( {{{\boldsymbol{y}}_{\rm{t}}}} \right)p\left( {{{\boldsymbol{y}}_{\boldsymbol{s}}}} \right)} \right]$ is intractable since both ${p\left( {{{\boldsymbol{y}}_{\rm{t}}},{{\boldsymbol{y}}_{\rm{s}}}} \right)}$ and ${p\left( {{{\boldsymbol{y}}_{\rm{t}}}} \right)p\left( {{{\boldsymbol{y}}_{\boldsymbol{s}}}} \right)}$ involve  mixtures with a large number of components and are intractable. Therefore, instead of the true value of $I\left( {{{\boldsymbol{y}}_{\rm{t}}};{{\boldsymbol{y}}_{\rm{s}}}} \right)$, we estimate $I\left( {{{\boldsymbol{y}}_{\rm{t}}};{{\boldsymbol{y}}_{\rm{s}}}} \right)$ and minimize the estimation. 

We first sample several ${\boldsymbol{y}}$. Denote $\tau \left( {{{\boldsymbol{y}}_{\rm{t}}},{{\boldsymbol{y}}_{\rm{s}}}} \right)$ as the probability that ${{{\boldsymbol{y}}_{\rm{t}}}}$ is interdependent with ${{{\boldsymbol{y}}_{\rm{s}}}}$. We have
\begin{equation}
\label{DIB-JSCC-C-1}
\tau \left( {{{\boldsymbol{y}}_{\rm{t}}},{{\boldsymbol{y}}_{\rm{s}}}} \right) = 
\begin{cases}
\begin{aligned}
0 &\quad {{\text{if }}p\left( {{{\boldsymbol{y}}_{\rm{t}}},{{\boldsymbol{y}}_{\rm{s}}}} \right) = p\left( {{{\boldsymbol{y}}_{\rm{t}}}} \right)p\left( {{{\boldsymbol{y}}_{\boldsymbol{s}}}} \right)}\\
1 &\quad {{\text{else}}}
\end{aligned}
\end{cases}.
\end{equation}
If ${{{\boldsymbol{y}}_{\rm{t}}}}$ and ${{{\boldsymbol{y}}_{\rm{s}}}}$ are sampled from ${p\left( {{{\boldsymbol{y}}_{\rm{t}}}} \right)p\left( {{{\boldsymbol{y}}_{\boldsymbol{s}}}} \right)}$, $\tau \left( {{{\boldsymbol{y}}_{\rm{t}}},{{\boldsymbol{y}}_{\rm{s}}}} \right) = 0$. If ${{{\boldsymbol{y}}_{\rm{t}}}}$ and ${{{\boldsymbol{y}}_{\rm{s}}}}$ are sampled from ${p\left( {{{\boldsymbol{y}}_{\rm{t}}},{{\boldsymbol{y}}_{\rm{s}}}} \right)}$, $\tau \left( {{{\boldsymbol{y}}_{\rm{t}}},{{\boldsymbol{y}}_{\rm{s}}}} \right) = 1$. Then, $I\left( {{{\boldsymbol{y}}_{\rm{t}}};{{\boldsymbol{y}}_{\rm{s}}}} \right)$ can be changed into\cite{pmlr-v80-kim18b, NEURIPS2018_1ee3dfcd, Chen_2021_ICCV}
\begin{equation}
\label{DIB-JSCC-C-2}
\begin{aligned}
I\left( {{{\boldsymbol{y}}_{\rm{t}}};{{\boldsymbol{y}}_{\rm{s}}}} \right) & = {\mathbb{E}_{p\left( {{{\boldsymbol{y}}_{\rm{t}}},{{\boldsymbol{y}}_{\rm{s}}}} \right)}}\left( {\log \frac{{p\left( {{{\boldsymbol{y}}_{\rm{t}}},{{\boldsymbol{y}}_{\rm{s}}}} \right)}}{{p\left( {{{\boldsymbol{y}}_{\rm{t}}}} \right)p\left( {{{\boldsymbol{y}}_{\boldsymbol{s}}}} \right)}}} \right)\\
&= {\mathbb{E}_{p\left( {{{\boldsymbol{y}}_{\rm{t}}},{{\boldsymbol{y}}_{\rm{s}}}} \right)}}\left( {\log \frac{{p\left( {\tau \left( {{{\boldsymbol{y}}_{\rm{t}}},{{\boldsymbol{y}}_{\rm{s}}}} \right) = 1} \right)}}{{p\left( {\tau \left( {{{\boldsymbol{y}}_{\rm{t}}},{{\boldsymbol{y}}_{\rm{s}}}} \right) = 0} \right)}}} \right).
\end{aligned}
\end{equation}
According to (\ref{DIB-JSCC-C-1}), $p\left( {\tau \left( {{{\boldsymbol{y}}_{\rm{t}}},{{\boldsymbol{y}}_{\rm{s}}}} \right) = 1} \right) + p\left( {\tau \left( {{{\boldsymbol{y}}_{\rm{t}}},{{\boldsymbol{y}}_{\rm{s}}}} \right) = 0} \right) = 1$. (\ref{DIB-JSCC-C-2}) can be changed into
\begin{equation}
\label{DIB-JSCC-C-3}
I\left( {{{\boldsymbol{y}}_{\rm{t}}};{{\boldsymbol{y}}_{\rm{s}}}} \right) = {\mathbb{E}_{p\left( {{{\boldsymbol{y}}_{\rm{t}}},{{\boldsymbol{y}}_{\rm{s}}}} \right)}}\left( {\log \frac{{p\left( {\tau \left( {{{\boldsymbol{y}}_{\rm{t}}},{{\boldsymbol{y}}_{\rm{s}}}} \right) = 1} \right)}}{{1 - p\left( {\tau \left( {{{\boldsymbol{y}}_{\rm{t}}},{{\boldsymbol{y}}_{\rm{s}}}} \right) = 1} \right)}}} \right).
\end{equation}

From (\ref{DIB-JSCC-C-3}), the estimation of $I\left( {{{\boldsymbol{y}}_{\rm{t}}};{{\boldsymbol{y}}_{\rm{s}}}} \right)$ requires only the probability ${p\left( {\tau \left( {{{\boldsymbol{y}}_{\rm{t}}},{{\boldsymbol{y}}_{\rm{s}}}} \right) = 1} \right)}$. However, directly estimating ${p\left( {\tau \left( {{{\boldsymbol{y}}_{\rm{t}}},{{\boldsymbol{y}}_{\rm{s}}}} \right) = 1} \right)}$ using Monte Carlo does
not work due to high dimensions of ${{{\boldsymbol{y}}_{\rm{t}}}}$ and ${{{\boldsymbol{y}}_{\rm{s}}}}$\cite{pmlr-v80-kim18b}. Hence, we employ the density-ratio trick \cite{nguyen2010estimating} that involves a discriminator to approximate ${p\left( {\tau \left( {{{\boldsymbol{y}}_{\rm{t}}},{{\boldsymbol{y}}_{\rm{s}}}} \right) = 1} \right)}$. Suppose we have a discriminator ${\rm{Di}}{{\rm{s}}_{\boldsymbol{\varepsilon }}}:{\mathbb{R}^{M}} \to {[0,1]^2}$ consists of neural network with parameters ${\boldsymbol{\varepsilon }}$. The output of the discriminator ${{\rm{Di}}{{\rm{s}}_{\boldsymbol{\varepsilon }}}\left( {{{\boldsymbol{y}}_{\rm{t}}},{{\boldsymbol{y}}_{\rm{s}}}} \right)}$ is treated as ${p\left( {\tau \left( {{{\boldsymbol{y}}_{\rm{t}}},{{\boldsymbol{y}}_{\rm{s}}}} \right) = 1} \right)}$. We obtains samples from ${p\left( {{{\boldsymbol{y}}_{\rm{t}}},{{\boldsymbol{y}}_{\rm{s}}}} \right)}$ by first choosing ${\boldsymbol{x}}$ uniformly at random and then sampling from $p\left( {{{\boldsymbol{y}}_{\rm{t}}}|{\boldsymbol{x}}} \right)$ and $p\left( {{{\boldsymbol{y}}_{\rm{s}}}|{\boldsymbol{x}}} \right)$. We can also sample from ${p\left( {{{\boldsymbol{y}}_{\rm{t}}}} \right)p\left( {{{\boldsymbol{y}}_{\boldsymbol{s}}}} \right)}$ by first sampling from ${p\left( {{{\boldsymbol{y}}_{\rm{t}}},{{\boldsymbol{y}}_{\rm{s}}}} \right)}$ and then permuting ${{{\boldsymbol{y}}_{\rm{t}}}}$ and ${{{\boldsymbol{y}}_{\rm{s}}}}$ along the batch axis. Having access to samples from both ${\tau \left( {{{\boldsymbol{y}}_{\rm{t}}},{{\boldsymbol{y}}_{\rm{s}}}} \right) = 1}$ and ${\tau \left( {{{\boldsymbol{y}}_{\rm{t}}},{{\boldsymbol{y}}_{\rm{s}}}} \right) = 0}$, we train the discriminator ${\rm{Di}}{{\rm{s}}_{\boldsymbol{\varepsilon }}}$ to distinguish samples from ${p\left( {{{\boldsymbol{y}}_{\rm{t}}},{{\boldsymbol{y}}_{\rm{s}}}} \right)}$ and ${p\left( {{{\boldsymbol{y}}_{\rm{t}}}} \right)p\left( {{{\boldsymbol{y}}_{\boldsymbol{s}}}} \right)}$. The loss function of the discriminator is
\begin{equation}
\label{DIB-JSCC-C-4}
\mathop {\min }\limits_{\boldsymbol{\varepsilon }} \log {\rm{Di}}{{\rm{s}}_{\boldsymbol{\varepsilon }}}\left( {{{\boldsymbol{y}}_{\rm{t}}},{{\boldsymbol{y}}_{\rm{s}}}} \right) + \log \left( {1 - {\rm{Di}}{{\rm{s}}_{\boldsymbol{\varepsilon }}}\left( {{{{\boldsymbol{\tilde y}}}_{\rm{t}}},{{{\boldsymbol{\tilde y}}}_{\rm{s}}}} \right)} \right),
\end{equation}
where ${{{{\boldsymbol{\tilde y}}}_{\rm{t}}}}$ and ${{{{\boldsymbol{\tilde y}}}_{\rm{s}}}}$ are the results by randomly permuting ${{{\boldsymbol{y}}_{\rm{t}}}}$ and ${{{\boldsymbol{y}}_{\rm{s}}}}$ along the batch axis, respectively. By optimizing (\ref{DIB-JSCC-C-4}), the output of ${\rm{Di}}{{\rm{s}}_{\boldsymbol{\varepsilon }}}$ will be forced to approach 0 when ${{{\boldsymbol{y}}_{\rm{t}}}}$ and ${{{\boldsymbol{y}}_{\rm{s}}}}$ are 
independent and approach 1 when ${{{\boldsymbol{y}}_{\rm{t}}}}$ and ${{{\boldsymbol{y}}_{\rm{s}}}}$ are dependent. Once the discriminator is properly trained, (\ref{DIB-JSCC-C-3}) can be expressed as
\begin{equation}
\label{DIB-JSCC-C-5}
I\left( {{{\boldsymbol{y}}_{\rm{t}}};{{\boldsymbol{y}}_{\rm{s}}}} \right) \approx {\mathbb{E}_{p\left( {{{\boldsymbol{y}}_{\rm{t}}},{{\boldsymbol{y}}_{\rm{s}}}} \right)}}\left( {\log \frac{{{\rm{Di}}{{\rm{s}}_{\boldsymbol{\varepsilon }}}\left( {{{\boldsymbol{y}}_{\rm{t}}},{{\boldsymbol{y}}_{\rm{s}}}} \right)}}{{1 - {\rm{Di}}{{\rm{s}}_{\boldsymbol{\varepsilon }}}\left( {{{\boldsymbol{y}}_{\rm{t}}},{{\boldsymbol{y}}_{\rm{s}}}} \right)}}} \right).
\end{equation}

Overall, by replacing $I\left( {{{\boldsymbol{y}}_{\rm{s}}};{\boldsymbol{s}}} \right)$ and $I\left( {{{\boldsymbol{y}}_{\rm{t}}};{{\boldsymbol{y}}_{\rm{s}}}} \right)$ in (\ref{DIB-JSCC-1}) with (\ref{DIB-JSCC-B-4}) and (\ref{DIB-JSCC-C-5}), respectively, we can calculate the DIB  objective for image transmission privacy-protective JSCC. However, we experimentally observe that when simultaneously training ${f_{{{\boldsymbol{\varphi }}_{\rm{s}}}}}$ and ${f_{{{\boldsymbol{\varphi }}_{\rm{t}}}}}$,  the neural networks will converge to a degenerated solution, where all information is encoded in ${{{\boldsymbol{y}}_{\rm{s}}}}$, whereas ${{{\boldsymbol{y}}_{\rm{t}}}}$ holds almost no information. To prevent this undesirable solution, we adopt a two-step training strategy\cite{Hadad2018two}. In the first step, ${f_{{{\boldsymbol{\varphi }}_{\rm{s}}}}}$ and the classifier ${{\rm{C}}_{\boldsymbol{\gamma }}}$ are jointly trained using (\ref{DIB-JSCC-B-4}) as loss function to extract ${{{\boldsymbol{y}}_{\rm{s}}}}$ that contains private information. In the second step, ${f_{{{\boldsymbol{\varphi }}_{\rm{s}}}}}$ is fixed. ${f_{{{\boldsymbol{\varphi }}_{\rm{t}}}}}$ and ${{\rm{D}}_{{{\boldsymbol{\theta }}_{\rm{B}}}}}$ are jointly trained using (\ref{DIB-JSCC-C-5}) as loss function, followed by alternating training with ${\rm{Di}}{{\rm{s}}_{\boldsymbol{\varepsilon }}}$ using (\ref{DIB-JSCC-C-4}) as loss function in order to enable ${{{\boldsymbol{y}}_{\rm{t}}}}$ to capture public information. By training ${f_{{{\boldsymbol{\varphi }}_{\rm{s}}}}}$  in the first step, and fixing the values of its parameters in the second step, ${f_{{{\boldsymbol{\varphi }}_{\rm{s}}}}}$ has a limited capacity since it ignores most of the public information and thus enabling ${f_{{{\boldsymbol{\varphi }}_{\rm{t}}}}}$ to extract public information. The whole training procedure of the DIB objective for JSCC is summarized in Algorithm \ref{algorithm:DIB}.

\begin{algorithm}[t]
\caption{Disentangled IB guided JSCC}
\begin{algorithmic}[1]
\renewcommand{\algorithmicrequire}{\textbf{Input:}}
\REQUIRE
Dataset ($\mathcal{X}$); SNR of ${\eta _{{\rm{AB}}}}\left(  \cdot  \right)$ ${\rm{SN}}{{\rm{R}}_{{\rm{AB}}}}$;  Hyperparameters ${\alpha}$ and ${\beta}$; Maximum Disentangled Epochs $E_{{\rm{max}}}^{{\rm{disten}}}$; Maximum Classification Epochs $E_{{\rm{max}}}^{{\rm{cls}}}$;.
\renewcommand{\algorithmicrequire}{\textbf{Output:}}
\REQUIRE
Learned ${{\rm{D}}_{{{\boldsymbol{\theta }}_{\rm{B}}}}}$, ${{\rm{D}}_{{{\boldsymbol{\theta }}_{\rm{B}}}}}$, ${f_{{{\boldsymbol{\varphi }}_{\rm{s}}}}}$, ${f_{{{\boldsymbol{\varphi }}_{\rm{t}}}}}$, ${\rm{Di}}{{\rm{s}}_{\boldsymbol{\varepsilon }}}$ and ${{\rm{C}}_{\boldsymbol{\gamma }}}$.
\STATE Initialize ${f_{{{\boldsymbol{\varphi }}_{\rm{s}}}}}$ and ${{\rm{C}}_{\boldsymbol{\gamma }}}$: ${{\boldsymbol{\varphi }}_{\rm{s}}},{\boldsymbol{\gamma }} \leftarrow {\boldsymbol{\varphi }}_{\rm{s}}^{\left( 0 \right)},{{\boldsymbol{\gamma }}^{\left( 0 \right)}}$.
\FOR{$t = 1,2, \ldots ,E_{{\rm{max}}}^{{\rm{cls}}}$}
    \STATE Sample $B$ samples from Dataset: $\left( {{\boldsymbol{x}},{\boldsymbol{s}}} \right) \sim p\left( {{\boldsymbol{x}},{\boldsymbol{s}}} \right)$;
    \STATE Calculate ${{\mathcal{L}}_{\rm{C}}}\left( {{{\boldsymbol{\varphi }}_{\rm{s}}},{\boldsymbol{\gamma }}} \right) = \sum\limits_{\boldsymbol{x}} {\log {{\rm{C}}_{\boldsymbol{\gamma }}}\left( {{f_{{{\boldsymbol{\varphi }}_{\rm{s}}}}}\left( {\boldsymbol{x}} \right)} \right)} $;  
    \STATE Update ${{\boldsymbol{\varphi }}_{\rm{s}}},{\boldsymbol{\gamma }} \leftarrow \arg \min {{\mathcal{L}}_{\rm{C}}}\left( {{{\boldsymbol{\varphi }}_{\rm{s}}},{\boldsymbol{\gamma }}} \right)$;
\ENDFOR
\STATE Initialize ${f_{{{\boldsymbol{\varphi }}_{\rm{t}}}}}$, ${{\rm{D}}_{{{\boldsymbol{\theta }}_{\rm{B}}}}}$ and ${\rm{Di}}{{\rm{s}}_{\boldsymbol{\varepsilon }}}$:  ${{\boldsymbol{\varphi }}_{\rm{t}}},{{\boldsymbol{\theta }}_{\rm{B}}},{\boldsymbol{\varepsilon }} \leftarrow {\boldsymbol{\varphi }}_{\rm{t}}^{\left( 0 \right)},{\boldsymbol{\theta }}_{\rm{B}}^{\left( 0 \right)},{{\boldsymbol{\varepsilon }}^{\left( 0 \right)}}$.
\FOR{$t = 1,2, \ldots ,E_{{\rm{max}}}^{{\rm{disten}}}$}  
    \STATE Calculate ${L_{\rm{B}}}\left( {{{\boldsymbol{\theta }}_{\rm{B}}}} \right) = \sum\limits_{\boldsymbol{x}} {{\rm{d}}\left( {{\boldsymbol{x}},{{{\boldsymbol{\hat x}}}^{\boldsymbol{p}}}} \right)} $
    
    ${L_{\rm{A}}}\left( {{{\boldsymbol{\varphi }}_{\rm{t}}}} \right) = \sum\limits_{\boldsymbol{x}} {\left( {{\rm{d}}\left( {{\boldsymbol{x}},{{{\boldsymbol{\hat x}}}^{\boldsymbol{p}}}} \right) + \alpha \log \frac{{{\rm{Di}}{{\rm{s}}_{\boldsymbol{\varepsilon }}}\left( {{{\boldsymbol{y}}_{\rm{t}}},{{\boldsymbol{y}}_{\rm{s}}}} \right)}}{{1 - {\rm{Di}}{{\rm{s}}_{\boldsymbol{\varepsilon }}}\left( {{{\boldsymbol{y}}_{\rm{t}}},{{\boldsymbol{y}}_{\rm{s}}}} \right)}}} \right)} $
    
    \STATE Update ${{\boldsymbol{\theta }}_{\rm{B}}} \leftarrow \arg \min {L_{\rm{B}}}\left( {{{\boldsymbol{\theta }}_{\rm{B}}}} \right)$;
    \STATE Update ${{\boldsymbol{\varphi }}_{\rm{t}}} \leftarrow \arg \min {L_{\rm{A}}}\left( {{{\boldsymbol{\varphi }}_{\rm{t}}}} \right)$;

    \STATE Calculate
    
    ${L_{{\rm{dis}}}}\left( {\boldsymbol{\varepsilon }} \right) = \sum\limits_{\boldsymbol{x}} {\log \left( {{\rm{Di}}{{\rm{s}}_{\boldsymbol{\varepsilon }}}\left( {{{\boldsymbol{y}}_{\rm{t}}},{{\boldsymbol{y}}_{\rm{s}}}} \right)\left( {1 - {\rm{Di}}{{\rm{s}}_{\boldsymbol{\varepsilon }}}\left( {{{{\boldsymbol{\tilde y}}}_{\rm{t}}},{{{\boldsymbol{\tilde y}}}_{\rm{s}}}} \right)} \right)} \right)}$;  

    \STATE Update ${\boldsymbol{\varepsilon }} \leftarrow \arg \min {L_{{\rm{dis}}}}\left( {\boldsymbol{\varepsilon }} \right)$;

\ENDFOR
\end{algorithmic}\label{algorithm:DIB}
\end{algorithm}

Even though the public subcodewords ${{{\boldsymbol{y}}_{\rm{t}}}}$ can be directly transmitted through the channel, the private subcodewords ${{{\boldsymbol{y}}_{\rm{s}}}}$ still needs protection. Directly deleting ${{\boldsymbol{y}}_{\rm{s}}}$ will severely damage the image transmission quality due to the loss of private information. Therefore, in section \ref{PP-algorithm}, we further propose a privacy-protective algorithm that mitigates privacy leakage at Eve and recovers the private information at Bob. 

\section{Privacy-protective algorithm}\label{PP-algorithm}
This section first proposes a password-based protection process that converts the private subcodewords into secure representations. Then, we introduce the optimization objective of the privacy-protective algorithm and derive its differentiable form. Next, we describe the whole training process of DIB-PPJSCC that combines the DIB objective and the password-based privacy-protective algorithm.

\subsection{Password-based privacy-protective process}\label{PP-algorithm-A} 
The password-based privacy-protective process is shown in the dashed box on the right side of Fig. \ref{fig:DIB-PPJSCC}. Assume Alice and Bob share a legitimate-user-specific password ${\boldsymbol{p}} \in {\mathbb{Z}}{^{{\rm{Len}}}}, 0{\rm{ }} < {\rm{ }}{\boldsymbol{p}}{\rm{ }} \le {\rm{ }}{p_{{\rm{level}}}}$, where ${\rm{Len}}$ is the length of ${\boldsymbol{p}}$, and ${p_{{\rm{level}}}}$ is a positive integer greater than $0$. To protect ${{\boldsymbol{y}}_{\rm{s}}}$ from being eavesdropped by Eve during transmission, an encryptor consist of fully-connected (FC) layer, ${{\rm{T}}_{\boldsymbol{\phi}} }:{\mathbb{R}^{\left( {{M_s} + {\rm{Len}}} \right)}} \to {\mathbb{R}^{{M_s}}}$ is applied after the encoder at Alice to encrypt ${{\boldsymbol{y}}_{\rm{s}}}$ into protected private subcodeword ${{{\boldsymbol{y}}_{\rm{s}}}}\in {\mathbb{R}^{{M_{\rm{s}}}}
}$. The raw private subcodeword ${\boldsymbol{y}}_{\rm{s}}^{\boldsymbol{p}}$ is first concatenated with $\boldsymbol{p}$ and then input into ${{\rm{T}}_{\boldsymbol{\phi}} }$ to obtain the protected subcodeword ${\boldsymbol{y}}_{\rm{s}}^{\boldsymbol{p}}$:
\begin{equation}
\label{PP-algorithm-A-1}
{\boldsymbol{y}}_{\rm{s}}^{\boldsymbol{p}} = {{\rm{T}}_{\boldsymbol{\phi}} }\left( {{\rm{concat}}\left[ {{{\boldsymbol{y}}_{\rm{s}}},{\boldsymbol{p}}} \right]} \right),
\end{equation}
Since ${{{\boldsymbol{y}}_{\rm{t}}}}$ contains little private information, we do no protection on ${{{\boldsymbol{y}}_{\rm{t}}}}$. ${{{\boldsymbol{y}}_{\rm{t}}}}$ and ${\boldsymbol{y}}_{\rm{s}}^{\boldsymbol{p}}$ are directly concatenated to obtain the protected codeword ${{\boldsymbol{y}}^{\boldsymbol{p}}} = {\rm{concat}}\left[ {{{\boldsymbol{y}}_{\rm{t}}},{\boldsymbol{y}}_{\rm{s}}^{\boldsymbol{p}}} \right] \in {\mathbb{R}^{M}}$, which is transmitted by Alice through the channel. The noisy protected codeword ${\boldsymbol{\hat y}}_{\rm{B}}^{\boldsymbol{p}}$ that Bob received is
\begin{equation}
\label{PP-algorithm-A-2}
\begin{aligned}
{\boldsymbol{\hat y}}_{\rm{B}}^{\boldsymbol{p}} & = {\eta _{{\rm{AB}}}}\left( {{{\boldsymbol{y}}^{{\boldsymbol{p}}}}} \right)\\
& = {\eta _{{\rm{AB}}}}\left( {{\rm{concat}}\left[ {{{\boldsymbol{y}}_{\rm{t}}},{\boldsymbol{y}}_{\rm{s}}^{\boldsymbol{p}}} \right]} \right)\\
& = {\rm{concat}}\left[ {{{{{\boldsymbol{\hat y}}}_{{\rm{t,B}}}}},{\boldsymbol{\hat y}}_{{\rm{s,B}}}^{\boldsymbol{p}}} \right],
\end{aligned}
\end{equation}
where ${{{\boldsymbol{\hat y}}}_{{\rm{t,B}}}} = {\eta _{{\rm{AB}}}}\left( {{{\boldsymbol{y}}_{\rm{t}}}} \right)$ is the noisy public subcodeword received at Bob, and ${\boldsymbol{\hat y}}_{{\rm{s,B}}}^{\boldsymbol{p}} = {\eta _{{\rm{AB}}}}\left( {{\boldsymbol{y}}_{\rm{s}}^{\boldsymbol{p}}} \right)$ is the noisy protected private subcodeword received at Bob. At Bob, a decryptor consists of FC layer, ${\rm{T}}_{\boldsymbol{\phi}} ^{ - 1}:{\mathbb{R}^{\left( {{M_s} + {\rm{Len}}} \right)}} \to {\mathbb{R}^{{M_s}}}$ is applied to recover the private information. Similarly, ${{\boldsymbol{\hat y}}_{{\rm{s,B}}}^{\boldsymbol{p}}}$ is first concatenated with ${\boldsymbol{p}}$ and then input into ${\rm{T}}_{\boldsymbol{\phi}} ^{ - 1}$. The recovered noisy private subcodeword at Bob, ${\boldsymbol{\bar y}}_{{\rm{s,B}}}^{\boldsymbol{p}} \in \mathbb{R}^{{M}_{\rm{s}}}$ is
\begin{equation}
\label{PP-algorithm-A-3}
{\boldsymbol{\bar y}}_{{\rm{s,B}}}^{\boldsymbol{p}} = {\rm{T}}_{\boldsymbol{\phi}} ^{ - 1}\left( {{\rm{concat}}\left[ {{\boldsymbol{\hat y}}_{{\rm{s,B}}}^{\boldsymbol{p}},{\boldsymbol{p}}} \right]} \right).
\end{equation}
${{{{\boldsymbol{\hat y}}}_{{\rm{t,B}}}}}$ is concatenated with ${\boldsymbol{\bar y}}_{{\rm{s,B}}}^{\boldsymbol{p}}$ to obtain the recovered codeword at Bob ${\boldsymbol{\bar y}}_{\rm{B}}^{\boldsymbol{p}} \in \mathbb{R}^{{M}}$
\begin{equation}
\label{PP-algorithm-A-4}
{\boldsymbol{\bar y}}_{\rm{B}}^{\boldsymbol{p}} = {\rm{T}}_{\boldsymbol{\phi}} ^{ - 1}\left( {{\rm{concat}}\left[ {{{{\boldsymbol{\hat y}}}_{{\rm{t,B}}}},{\boldsymbol{\bar y}}_{{\rm{s,B}}}^{\boldsymbol{p}}} \right]} \right).
\end{equation}
According to (\ref{model-3}), ${\boldsymbol{\bar y}}_{\rm{B}}^{\boldsymbol{p}}$ is then input into the decoder to reconstruct the input image as
\begin{equation}
\label{SIP5}
{{{\boldsymbol{\hat x}}}^{\boldsymbol{p}}}{\rm{ = }}{{\rm{D}}_{{{\boldsymbol{\theta }}_{\rm{B}}}}}\left( {\boldsymbol{\bar y}}_{\rm{B}}^{\boldsymbol{p}} \right).
\end{equation}

Meanwhile, the adversarial eavesdropper Eve has access to the protected codeword ${{\boldsymbol{y}}^{\boldsymbol{p}}}$ through the channel between Alice and Eve. The noisy protected codeword received at Eve ${\boldsymbol{\hat y}}_{\rm{E}}^{\boldsymbol{p}}$ is
\begin{equation}
\label{PP-algorithm-A-5}
\begin{aligned}
{\boldsymbol{\hat y}}_{\rm{E}}^{\boldsymbol{p}} & = {\eta _{{\rm{AE}}}}\left( {{{\boldsymbol{y}}^{\rm{p}}}} \right)\\
& = {\eta _{{\rm{AE}}}}\left( {{\rm{concat}}\left[ {{{\boldsymbol{y}}_{\rm{t}}},{\boldsymbol{y}}_{\rm{s}}^{\boldsymbol{p}}} \right]} \right)\\
& = {\rm{concat}}\left[ {{{{\boldsymbol{\hat y}}}_{{\rm{t,E}}}},{\boldsymbol{\hat y}}_{{\rm{s,E}}}^{\boldsymbol{p}}} \right],    
\end{aligned}
\end{equation}
where ${{{\boldsymbol{\hat y}}}_{{\rm{t,E}}}} = {\eta _{{\rm{AE}}}}\left( {{{\boldsymbol{y}}_{\rm{t}}}} \right) \in {\mathbb{R}^{{M_{\rm{t}}}}} $ is the noisy public subcodeword received at Eve, and ${\boldsymbol{\hat y}}_{{\rm{s,E}}}^{\boldsymbol{p}} = {\eta _{{\rm{AE}}}}\left( {{\boldsymbol{y}}_{\rm{s}}^{\boldsymbol{p}}} \right) \in {\mathbb{R}^{{M_{\rm{s}}}}}$ is the noisy protected private subcodeword received at Eve. Eve will try to estimate $\boldsymbol{s}$ from ${\boldsymbol{\hat y}}_{\rm{E}}^{\boldsymbol{p}}$. The estimation is 
\begin{equation}
\label{PP-algorithm-A-6}
{\boldsymbol{\hat s}}_{\rm{E}}^{\boldsymbol{p}} = {{\rm{D}}_{{{\boldsymbol{\theta }}_{\rm{E}}}}}\left( {{\boldsymbol{\hat y}}_{\rm{E}}^{\boldsymbol{p}}} \right).
\end{equation}
In addition, we consider a severe scenario where Eve has access to the structure and parameters of ${\rm{T}}_{\boldsymbol{\phi}} ^{ - 1}\left(  \cdot  \right)$ as well as the structure of password, but has no access to the actual correct password $\boldsymbol{p}$. In this situation, Eve randomly guesses the password $\boldsymbol{p}$ shared by Alice and Bob and tries to recover the raw private subcodword ${{{\boldsymbol{y}}_{\rm{s}}}}$. Assume the password guessed by Eve is ${{{\boldsymbol{p}}_1}}$. The recovered noisy private subcodeword at Eve, ${\boldsymbol{\bar y}}_{{\rm{s,E}}}^{{\boldsymbol{p}},{{\boldsymbol{p}}_1}} \in {\mathbb{R}^{{M_{\rm{s}}}}}$ is
\begin{equation}
\label{PP-algorithm-A-7}
{\boldsymbol{\bar y}}_{{\rm{s,E}}}^{{\boldsymbol{p}},{{\boldsymbol{p}}_1}} = {\rm{T}}_{\boldsymbol{\phi}} ^{ - 1}\left( {{\rm{concat}}\left[ {{\boldsymbol{\hat y}}_{{\rm{s,E}}}^{\boldsymbol{p}},{{\boldsymbol{p}}_1}} \right]} \right).
\end{equation}
Denote the codeword that Eve uses to estimate $\boldsymbol{s}$ as ${\boldsymbol{\bar y}}_{\rm{E}}^{{\boldsymbol{p}},{{\boldsymbol{p}}_1}}{\rm{ = concat}}\left[ {{{{\boldsymbol{\hat y}}}_{{\rm{t,E}}}},{\boldsymbol{\bar y}}_{{\rm{s,E}}}^{{\boldsymbol{p}},{{\boldsymbol{p}}_1}}} \right]$. The private information estimated by Eve in this situation is
\begin{equation}
\label{PP-algorithm-A-8}
{\boldsymbol{\bar s}}_{\rm{E}}^{{\boldsymbol{p}},{{\boldsymbol{p}}_1}} = {{{\rm{\bar D}}}_{{{{\boldsymbol{\bar \theta }}}_{\rm{E}}}}}\left( {{\boldsymbol{\bar y}}_{\rm{E}}^{{\boldsymbol{p}},{{\boldsymbol{p}}_1}}} \right),
\end{equation}
where ${{{\rm{\bar D}}}_{{{{\boldsymbol{\bar \theta }}}_{\rm{E}}}}}:{\mathbb{R}^M} \to {\left\{ {0,1} \right\}^S}$ is neural networks with parameters ${{{{\boldsymbol{\bar \theta }}}_{\rm{E}}}}$ and structure similar to that of ${{\rm{D}}_{{{\boldsymbol{\theta }}_{\rm{E}}}}}\left(  \cdot  \right)$. The goal of the password-based privacy-protective process is to minimize the amount of private information in ${\boldsymbol{\hat y}}_{\rm{E}}^{\boldsymbol{p}}$ and ${\boldsymbol{\bar y}}_{\rm{E}}^{{\boldsymbol{p}},{{\boldsymbol{p}}_1}}$ while maximizing the amount of private information in ${\boldsymbol{\bar y}}_{{\rm{s,B}}}^{\boldsymbol{p}}$. This ensures that Bob can successfully recover all the necessary private and public information, while Eve remains incapable of eavesdropping the private information.

\begin{algorithm}[t]
\caption{PP-algorithm}
\begin{algorithmic}[1]
\renewcommand{\algorithmicrequire}{\textbf{Input:}}
\REQUIRE
Dataset ($\mathcal{X}$); SNR of ${\eta _{{\rm{AB}}}}$ ${\rm{SN}}{{\rm{R}}_{{\rm{AB}}}}$; SNR of ${\eta _{{\rm{AE}}}}$ ${\rm{SN}}{{\rm{R}}_{{\rm{AE}}}}$; Hyperparameters ${\alpha _1}$ and ${\beta _1}$; Trained ${f_{{{\boldsymbol{\varphi }}_{\rm{s}}}}}\left(  \cdot  \right)$ and ${f_{{{\boldsymbol{\varphi }}_{\rm{t}}}}}\left(  \cdot  \right)$, Maximum Epochs ${E_{\max }}$; Password level ${{p_{{\rm{level}}}}}$.
\renewcommand{\algorithmicrequire}{\textbf{Output:}}
\REQUIRE
Learned ${{\rm{D}}_{{{\boldsymbol{\theta }}_{\rm{B}}}}}$, ${{\rm{T}}_{\boldsymbol{\phi}} }$, ${\rm{T}}_{\boldsymbol{\phi}} ^{ - 1}$, ${{\rm{D}}_{{{\boldsymbol{\theta }}_{\boldsymbol{E}}}}}$ and ${{{\rm{\bar D}}}_{{{{\boldsymbol{\bar \theta }}}_{\rm{E}}}}}$.
\STATE Initialize ${{\rm{D}}_{{{\boldsymbol{\theta }}_{\rm{B}}}}}$, ${{\rm{T}}_{\boldsymbol{\phi}} }$, ${\rm{T}}_{\boldsymbol{\phi}} ^{ - 1}$, ${{\rm{D}}_{{{\boldsymbol{\theta }}_{\boldsymbol{E}}}}}$ and ${{{\rm{\bar D}}}_{{{{\boldsymbol{\bar \theta }}}_{\rm{E}}}}}$: 

${{\boldsymbol{\theta }}_{\rm{B}}},{\boldsymbol{\phi}} ,{{\boldsymbol{\theta }}_{\rm{E}}},{{{\boldsymbol{\bar \theta }}}_{\rm{E}}} \leftarrow {\boldsymbol{\theta }}_{\rm{B}}^{\left( 0 \right)},{{\boldsymbol{\phi}} ^{\left( 0 \right)}},{\boldsymbol{\theta }}_{\rm{E}}^{\left( 0 \right)},{\boldsymbol{\bar \theta }}_{\rm{E}}^{\left( 0 \right)}$.
\FOR{$t = 1,2, \ldots ,{E_{\max }}$}
  \STATE Sample $B$ samples from Dataset: $\left( {{\boldsymbol{x}},{\boldsymbol{s}}} \right) \sim p\left( {{\boldsymbol{x}},{\boldsymbol{s}}} \right)$;
  \STATE Sample $B$ passwords ${\boldsymbol{p}} \sim {\rm{U}}\left( {1,{p_{{\rm{level}}}}} \right)$;
  \STATE Sample $B$ passwords ${{\boldsymbol{p}}_1} \sim {\rm{U}}\left( {1,{p_{{\rm{level}}}}} \right)$ guessed by Eve;
  \STATE Calculate ${{\mathcal{L}}_{\rm{B}}}\left( {{{\boldsymbol{\theta }}_{\rm{B}}}} \right){\rm{ = }}\sum\limits_{\boldsymbol{x}} {{\rm{d}}\left( {{\boldsymbol{x}},{{{\boldsymbol{\hat x}}}^{\boldsymbol{p}}}} \right)} $, 
  
  ${{\mathcal{L}}_{\rm{T}}}\left( {\boldsymbol{\phi}}  \right){\rm{ = }}\sum\limits_{\boldsymbol{x}} {{\rm{d}}\left( {{\boldsymbol{x}},{{{\boldsymbol{\hat x}}}^{\boldsymbol{p}}}} \right) - {\alpha _1}H\left( {{\boldsymbol{\hat s}}} \right) - {\beta _1}H\left( {{\boldsymbol{\bar s}}_{\rm{E}}^{{\boldsymbol{p}},{{\boldsymbol{p}}_1}}} \right)} $, 
  
  ${{\mathcal{L}}_{\rm{E}}}\left( {{{\boldsymbol{\theta }}_{\rm{E}}},{{{\boldsymbol{\bar \theta }}}_{\rm{E}}}} \right) = \sum\limits_{\boldsymbol{x}} {\log \left( {{{\rm{D}}_{{{\boldsymbol{\theta }}_{\boldsymbol{E}}}}}\left( {{\boldsymbol{\hat y}}_{\rm{E}}^{\boldsymbol{p}}} \right){{{\rm{\bar D}}}_{{{{\boldsymbol{\bar \theta }}}_{\rm{E}}}}}\left( {{\boldsymbol{\bar y}}_{\rm{E}}^{{\boldsymbol{p}},{{\boldsymbol{p}}_1}}} \right)} \right)} $;
  \STATE Update ${{\boldsymbol{\theta }}_{\rm{E}}},{{{\boldsymbol{\bar \theta }}}_{\rm{E}}} \leftarrow \arg \min {{\mathcal{L}}_{\rm{E}}}\left( {{{\boldsymbol{\theta }}_{\rm{E}}},{{{\boldsymbol{\bar \theta }}}_{\rm{E}}}} \right)$;
  \STATE Update ${{\boldsymbol{\theta }}_{\rm{B}}},{\boldsymbol{\phi}}  \leftarrow \arg \min \left( {{{\mathcal{L}}_{\rm{B}}}\left( {{{\boldsymbol{\theta }}_{\rm{B}}}} \right) + {{\mathcal{L}}_{\rm{T}}}\left( {\boldsymbol{\phi}}  \right)} \right)$;
\ENDFOR
\end{algorithmic}\label{algorithm: PP}
\end{algorithm}

\subsection{Loss function}
To guarantee the image transmission quality as well as the privacy-protection effectiveness, we design an optimization objective for the aforementioned protection process as
\begin{equation}
\label{PP-algorithm-B-1}
\begin{aligned}
&\mathop {{\rm{min}}}\limits_{{{\boldsymbol{\theta }}_{\rm{B}}}} {{\mathbb{E}}_{p\left( {{\boldsymbol{x,s}}} \right)}}\left( {{\rm{d}}\left( {{\boldsymbol{x}},{{{\boldsymbol{\hat x}}}^{\boldsymbol{p}}}} \right)} \right)\\
&\mathop {{\rm{min}}}\limits_{\boldsymbol{\phi}}  {{\mathbb{E}}_{p\left( {{\boldsymbol{x,s}}} \right)}}\left( {{\rm{d}}\left( {{\boldsymbol{x}},{{{\boldsymbol{\hat x}}}^{\boldsymbol{p}}}} \right)} \right) + {\alpha _1}{D_{{\rm{KL}}}}\left( {{\boldsymbol{\hat s}}||{\rm{U}}} \right)\\ 
&\quad\quad + {\beta _1}{D_{{\rm{KL}}}}\left( {{\boldsymbol{\bar s}}_{\rm{E}}^{{\boldsymbol{p}},{{\boldsymbol{p}}_1}}||{\rm{U}}} \right)\\
&\mathop {{\rm{max}}}\limits_{{{\boldsymbol{\theta }}_{\rm{E}}},{{{\boldsymbol{\bar \theta }}}_{\rm{E}}}} I\left( {{\boldsymbol{\hat y}}_{\rm{E}}^{\boldsymbol{p}},{\boldsymbol{s}}} \right) + I\left( {{\boldsymbol{\bar y}}_{\rm{E}}^{{\boldsymbol{p}},{{\boldsymbol{p}}_1}},{\boldsymbol{s}}} \right)
\end{aligned}
\end{equation}
where ${\alpha _1}$ and ${\beta _1}$ are hyperparameters, $\rm{U}$ is the uniform distribution. The optimization objective comprises three terms associated with different components: the decoder at Bob ${{\rm{D}}_{{{\boldsymbol{\theta }}_{\rm{B}}}}}$, the encryptor ${{\rm{T}}_{\boldsymbol{\phi}} }$ and the decryptor ${\rm{T}}_{\boldsymbol{\phi}} ^{ - 1}$, the private information estimator ${{\rm{D}}_{{{\boldsymbol{\theta }}_{\boldsymbol{E}}}}}$ and ${{{\rm{\bar D}}}_{{{{\boldsymbol{\bar \theta }}}_{\rm{E}}}}}$ at Eve. The first line in (\ref{PP-algorithm-B-1}) optimizes ${{\rm{D}}_{{{\boldsymbol{\theta }}_{\rm{B}}}}}$ to minimize the reconstruction distortion between the raw image $\boldsymbol{x}$ and the reconstructed image ${{{{\boldsymbol{\hat x}}}^{\boldsymbol{p}}}}$ at Bob. This term enables the enhancement of the reconstruction quality. The second line in (\ref{PP-algorithm-B-1}) also optimizes ${{\rm{T}}_{\boldsymbol{\phi}} }$ and ${\rm{T}}_{\boldsymbol{\phi}} ^{ - 1}$ to reduce the reconstruction distortion so as to ensure reconstruction quality while protecting private information.  Additionally, this line optimizes ${{\rm{T}}_{\boldsymbol{\phi}} }$ and ${\rm{T}}_{\boldsymbol{\phi}} ^{ - 1}$ to minimize the KL divergence between ${{\boldsymbol{\hat s}}}$ and uniform distribution, as well as the KL divergence between ${{\boldsymbol{\bar s}}_{\rm{E}}^{{\boldsymbol{p}},{{\boldsymbol{p}}_1}}}$ and the uniform distribution. By doing so, the estimations based on ${{\boldsymbol{\hat y}}_{\rm{E}}^{\boldsymbol{p}}}$ and ${{\boldsymbol{\bar y}}_{\rm{E}}^{{\boldsymbol{p}},{{\boldsymbol{p}}_1}}}$ become more randomized, and ${{\boldsymbol{\hat y}}_{\rm{E}}^{\boldsymbol{p}}}$ and ${{\boldsymbol{\bar y}}_{\rm{E}}^{{\boldsymbol{p}},{{\boldsymbol{p}}_1}}}$ contain minimal information about ${\boldsymbol{s}}$ \cite{Roy_2019_CVPR}. According to the definition of KL divergence, ${D_{{\rm{KL}}}}\left( {{\boldsymbol{\hat s}}||{\rm{U}}} \right)$ and ${D_{{\rm{KL}}}}\left( {{\boldsymbol{\bar s}}_{\rm{E}}^{{\boldsymbol{p}},{{\boldsymbol{p}}_1}}||{\rm{U}}} \right)$ are calculated as
\begin{equation}
\label{SIP-loss2}
\begin{aligned}
{D_{{\rm{KL}}}}\left( {{\boldsymbol{\hat s}}||{\rm{U}}} \right)& = {\mathbb{E}_{p\left( {{\boldsymbol{\hat s}}} \right)}}\left( {\log \left( {\frac{{p\left( {{\boldsymbol{\hat s}}} \right)}}{{p\left( {\rm{U}} \right)}}} \right)} \right)\\
& = \log S - H\left( {{\boldsymbol{\hat s}}} \right)\\
{D_{{\rm{KL}}}}\left( {{\boldsymbol{\bar s}}_{\rm{E}}^{{\boldsymbol{p}},{{\boldsymbol{p}}_1}}||{\rm{U}}} \right) & = {\mathbb{E}_{p\left( {{\boldsymbol{\bar s}}_{\rm{E}}^{{\boldsymbol{p}},{{\boldsymbol{p}}_1}}} \right)}}\left( {\log \left( {\frac{{p\left( {{\boldsymbol{\bar s}}_{\rm{E}}^{{\boldsymbol{p}},{{\boldsymbol{p}}_1}}} \right)}}{{p\left( {\rm{U}} \right)}}} \right)} \right)\\
& = \log S - H\left( {{\boldsymbol{\bar s}}_{\rm{E}}^{{\boldsymbol{p}},{{\boldsymbol{p}}_1}}} \right).
\end{aligned}
\end{equation}
Since $S$ is a fixed constant that cannot be optimized, minimizing ${D_{{\rm{KL}}}}\left( {{\boldsymbol{\hat s}}||{\rm{U}}} \right)$ and ${D_{{\rm{KL}}}}\left( {{\boldsymbol{\bar s}}_{\rm{E}}^{{\boldsymbol{p}},{{\boldsymbol{p}}_1}}||{\rm{U}}} \right)$ is equal to maximizing the entropy of ${{\boldsymbol{\hat s}}}$ and ${{\boldsymbol{\bar s}}_{\rm{E}}^{{\boldsymbol{p}},{{\boldsymbol{p}}_1}}}$, thus increasing the uncertainty of the private estimations made by Eve. The third line in (\ref{PP-algorithm-B-1}) optimizes ${{\rm{D}}_{{{\boldsymbol{\theta }}_{\boldsymbol{E}}}}}$ and ${{{\rm{\bar D}}}_{{{{\boldsymbol{\bar \theta }}}_{\rm{E}}}}}$ at Eve to maximizes $I\left( {{\boldsymbol{\hat y}}_{\rm{E}}^{\boldsymbol{p}},{\boldsymbol{s}}} \right)$ and $I\left( {{\boldsymbol{\bar y}}_{\rm{E}}^{{\boldsymbol{p}},{{\boldsymbol{p}}_1}},{\boldsymbol{s}}} \right)$. This optimization goal aims to improve Eve's eavesdropping accuracy on private information ${\boldsymbol{s}}$. When Alice and Bob engage in alternating training with Eve, they can obtain stronger privacy protection performance. However, the mutual information used in (\ref{PP-algorithm-B-1}) cannot be calculated due to the unknown $p\left( {{\boldsymbol{\hat y}}_{\rm{E}}^{\boldsymbol{p}}} \right)$ and $p\left( {{\boldsymbol{\bar y}}_{\rm{E}}^{{\boldsymbol{p}},{{\boldsymbol{p}}_1}}} \right)$. We use ${{\rm{D}}_{{{\boldsymbol{\theta }}_{\rm{E}}}}}$ and ${{{\rm{\bar D}}}_{{{{\boldsymbol{\bar \theta }}}_{\rm{E}}}}}$ to estimate $I\left( {{\boldsymbol{\hat y}}_{\rm{E}}^{\boldsymbol{p}},{\boldsymbol{s}}} \right)$ and $I\left( {{\boldsymbol{\bar y}}_{\rm{E}}^{{\boldsymbol{p}},{{\boldsymbol{p}}_1}},{\boldsymbol{s}}} \right)$. Let $q\left( {{\boldsymbol{s}}|{\boldsymbol{\hat y}}_{\rm{E}}^{\boldsymbol{p}}} \right) = {{\rm{D}}_{{{\boldsymbol{\theta }}_{\boldsymbol{E}}}}}\left( {{\boldsymbol{\hat y}}_{\rm{E}}^{\boldsymbol{p}}} \right)$ and $q\left( {{\boldsymbol{s}}|{\boldsymbol{\bar y}}_{\rm{E}}^{{\boldsymbol{p}},{{\boldsymbol{p}}_1}}} \right) = {{{\rm{\bar D}}}_{{{{\boldsymbol{\bar \theta }}}_{\rm{E}}}}}\left( {{\boldsymbol{\bar y}}_{\rm{E}}^{{\boldsymbol{p}},{{\boldsymbol{p}}_1}}} \right)$. Similar to (\ref{DIB-JSCC-B-4}), we have
\begin{equation}
\label{SIP-loss3}
\begin{aligned}
& I\left( {{\boldsymbol{\hat y}}_{\rm{E}}^{\boldsymbol{p}},{\boldsymbol{s}}} \right) \ge {{\mathbb{E}}_{p\left( {{\boldsymbol{x,s}}} \right)}}\left( {\log {{\rm{D}}_{{{\boldsymbol{\theta }}_{\boldsymbol{E}}}}}\left( {{\boldsymbol{\hat y}}_{\rm{E}}^{\boldsymbol{p}}} \right)} \right)\\
& I\left( {{\boldsymbol{\bar y}}_{\rm{E}}^{{\boldsymbol{p}},{{\boldsymbol{p}}_1}},{\boldsymbol{s}}} \right) \ge {{\mathbb{E}}_{p\left( {{\boldsymbol{x,s}}} \right)}}\left( {\log {{{\rm{\bar D}}}_{{{{\boldsymbol{\bar \theta }}}_{\rm{E}}}}}\left( {{\boldsymbol{\bar y}}_{\rm{E}}^{{\boldsymbol{p}},{{\boldsymbol{p}}_1}}} \right)} \right).
\end{aligned}
\end{equation}
By substituting (\ref{SIP-loss2}) and (\ref{SIP-loss3}) into (\ref{PP-algorithm-B-1}), we can derive the differentiable form of the proposed loss function. The training procedure of PP algorithm is summarized in Algorithm \ref{algorithm: PP}. 

\subsection{Training Process of DIB-PPJSCC}
The architecture of the DIB-PPJSCC system is shown in Fig. \ref{fig:DIB-PPJSCC}. The encoder first extracts the information of the input image as features according to feature extractors. Then, to control the length of ${{{\boldsymbol{y}}_{\rm{t}}}}$ and ${{{\boldsymbol{y}}_{\rm{s}}}}$, FC layers are used to turn the features into vectors ${{{\boldsymbol{y}}_{\rm{t}}}}$ and ${{{\boldsymbol{y}}_{\rm{s}}}}$. At Bob, ${{{\boldsymbol{\hat y}}}_{\rm{B}}}$ is first passed into an FC layer and then upsampled to the same dimension as $\boldsymbol{x}$ to obtain ${{\boldsymbol{\hat x}}}$. At the output layer of ${{\rm{D}}_{{{\boldsymbol{\theta }}_{\rm{B}}}}}$, an activation function is used to transform the pixel values in ${{\boldsymbol{\hat x}}}$ to $\left[ {0,1} \right]$. We then multiply ${{\boldsymbol{\hat x}}}$ by 255 and round the resulting values to ensure that the pixel values are discrete and fall between $\left[ {0,255} \right]$. At Eve, ${{{\boldsymbol{\hat y}}}_{\rm{E}}}$ is directly input into ${{\rm{D}}_{{{\boldsymbol{\theta }}_{\rm{E}}}}}$ and ${{{\rm{\bar D}}}_{{{{\boldsymbol{\bar \theta }}}_{\rm{E}}}}}$. The softmax layer\cite{Hin2006Fast} is 
applied to calculate the estimated probabilities at the output layers of ${{\rm{D}}_{{{\boldsymbol{\theta }}_{\rm{E}}}}}$ and ${{{\rm{\bar D}}}_{{{{\boldsymbol{\bar \theta }}}_{\rm{E}}}}}$. The whole training process of DIB-PPJSCC consists of two stages. In the first stage, the encoder ${{\rm{E}}_{\boldsymbol{\varphi }}}$ and the classifier ${{\rm{C}}_{\boldsymbol{\gamma }}}$ are trained to disentangle ${{{\boldsymbol{y}}_{\rm{t}}}}$ and ${{{\boldsymbol{y}}_{\rm{s}}}}$ following Algorithm  \ref{algorithm:DIB}. In particular, ${f_{{{\boldsymbol{\varphi }}_{\rm{s}}}}}$ and ${{\rm{C}}_{\boldsymbol{\gamma }}}$ are jointly trained to improve the classification accuracy of ${{{\boldsymbol{y}}_{\rm{s}}}}$ for ${V_{{{\rm{d}}_1}}}$ epochs, and then ${f_{{{\boldsymbol{\varphi }}_{\rm{t}}}}}$ is trained to extract ${{{\boldsymbol{y}}_{\rm{t}}}}$ for ${V_{{{\rm{d}}_2}}}$ epochs. In the second stage,  ${f_{{{\boldsymbol{\varphi }}_{\rm{s}}}}}$ and ${f_{{{\boldsymbol{\varphi }}_{\rm{t}}}}}$ are fixed. ${{\rm{D}}_{{{\boldsymbol{\theta }}_{\rm{B}}}}}$, ${{\rm{T}}_\phi }$, ${\rm{T}}_\phi ^{ - 1}$ ${{\rm{D}}_{{{\boldsymbol{\theta }}_{\rm{E}}}}}$ and ${{{\rm{\bar D}}}_{{{{\boldsymbol{\bar \theta }}}_{\rm{E}}}}}$ are jointly trained using (\ref{PP-algorithm-B-1}) as loss function for ${V_{\rm{p}}}$ epochs to achieve the password-based privacy-protection.


\begin{table*}[t!]
\setlength\tabcolsep{5pt}
\renewcommand\arraystretch{1.1}
\centering
\caption{\label{tab:Acc} Eavesdropping accuracy under different ${\rm{SN}}{{\rm{R}}_{{\rm{AE}}}}$.}
\begin{tabular}{c|c|ccccccc}
\toprule
\multirow{2}*{\textbf{Datasets}} & \multirow{2}*{\textbf{Methods}} & \multicolumn{7}{c}{\textbf{Acc under different ${\textbf{SN}}{{\textbf{R}}_{{\textbf{AE}}}}$} } \\
~ & ~ & -15dB & -10dB & -5dB & 0dB & 5dB & 10dB & 15dB\\
\midrule
\multirow{4}*{Colored MNIST} & Deep JSCC + randomly discard & 0.125 & 0.1797 & 0.2656 & 0.4141 & 0.4297 & 0.5313 & 0.5703 \\   
~ &  DIB-JSCC + private codewords discard & \textbf{0.0781} & 0.1328 & \textbf{0.1093} & \textbf{0.125} & \textbf{0.1172} & \textbf{0.1172} & \textbf{0.1406} \\
~ & Adversarial JSCC & 0.125 & 0.1328 & 0.1849 & 0.2969 & 0.3125 &  0.3672 & 0.3647 \\
~ & DIB-PPJSCC & 0.0938 & \textbf{0.1171} &  0.1249 & 0.1302 & 0.1406 & 0.1823 & 0.1953\\
\midrule
\multirow{4}*{UTK face}  & Deep JSCC + randomly discard  & 0.2968 & 0.3593 & 0.4063 & 0.5469 & 0.5938 & 0.6093 & 0.6094 \\ 
~ & DIB-JSCC + private codewords discard & 0.2344 & 0.2657 & 0.375 & 0.3906 & 0.4063 & 0.4375 & 0.4375 \\
~ & Adversarial JSCC& 0.2344 & 0.3438 & 0.3906 & 0.4063 & 0.4063 & 0.4375 & 0.4844 \\
~ & DIB-PPJSCC & \textbf{0.2188} & \textbf{0.25} & \textbf{0.3281} & \textbf{0.375} & \textbf{0.375} & \textbf{0.3906} & \textbf{0.4063} \\
\bottomrule
\end{tabular}
\end{table*}
\section{Experimental Results}\label{PP-algorithm}
In this section, we provide extensive experiments to validate DIB-PPJSCC. We compare the image transmission and privacy protection performance of DIB-PPJSCC with various baselines by assessing the reconstruction error and privacy eavesdropping accuracy as key metrics. Furthermore, we employed visualizations of the extracted codewords and recovered images to illustrate the disentanglement and privacy protection capabilities of DIB-PPJSCC, and the robustness and complexity of DIB-PPJSCC are also discussed.  

\subsection{Experimental settings}\label{Experimental settings}
\subsubsection{Datasets}The experiments are carried on two datasets including the colored MNIST\cite{lecun1998mnist} and the UTK Face dataset\cite{Zhang2017Age} to account for different image sizes and colors. 


\textbf{The colored MNIST dataset:} The raw MNIST dataset comprises grayscale images of handwritten digits ranging from ``0" to ``9", consisting of a train set with 60000 samples and a test set with 10000 samples. We extend it into the colored MNIST dataset by randomly and uniformly assigning ten different colors to the handwritten digits. The generated colored MNIST dataset is RGB, and we set the color of the handwritten digits as the private information ${\boldsymbol{s}}$. The raw images are normalized to the range of $\left[ {0,1} \right]$ by dividing $255$.

\textbf{The UTK Face dataset:} It has a total of 23705 facial images of individuals from different age groups, ethnicities, and genders without
the specification of the train and test dataset. We set ethnicity to the private information ${\boldsymbol{s}}$. The dataset provides annotations for five categories of ethnicity: White, Black, Asian, Indian, and an "Other" category including individuals from various other ethnic backgrounds. We randomly select 19000 images for the train set, while the remaining images are allocated to the test set. The raw images sized at $200 \times 200 \times 3$, are first cropped into $128 \times 128 \times 3$ and then normalized to the range of $\left[ {0,1} \right]$ by dividing $255$.

\subsubsection{Baselines} We compare DIB-PPJSCC with the following four baselines to illustrate the advantages of different components of DIB-PPJSCC over existing studies. 

\textbf{A. Deep JSCC\cite{bourtsoulatze2019deep} + randomly discard:} The model consists of a JSCC encoder and decoder that are jointly trained to reduce the image transmission error without considering privacy protection. During testing, part of the codewords extracted by the JSCC encoder is randomly discarded, while the remaining codewords are transmitted.

\textbf{B. DIB-JSCC + private codewords discard:} The JSCC encoder and decoder in this method are trained to reduce the DIB objective following Algorithm \ref{algorithm:DIB}. After training, the extracted codewords consist of private codewords and public codewords. During testing, this model only transmits the public codewords and does not transmit the private codewords.

\textbf{C. Adversarial JSCC\cite{Marchioro2020Adversarial, Erdemir2022Privacy}:} This model is composed of a JSCC encoder followed by a JSCC decoder at Bob for image reconstruction and a classifier at Eve for private information classification. The JSCC encoder and decoder are jointly trained to reduce the reconstruction error as well as increase the uncertainty of privacy classification at Eve. Meanwhile, a classifier at Eve is trained to improve the eavesdropping accuracy on the private information.

\textbf{D. Separate scheme:} This baseline employs separate source coding, encryption, and channel coding followed by modulation. We choose the better portable graphics (BPG)\cite{BPG} for source coding, the advanced encryption standard (AES)\cite{AES} with block and key
size of 128 bits for encryption, the low-density parity-check (LDPC)\cite{LDPC} codes with $\frac{1}{2}$ rate for channel coding, and the 4-ary quadrature amplitude modulation (QAM) for modulation.

\begin{figure}[t]
\centering{
\includegraphics[width=1\linewidth]{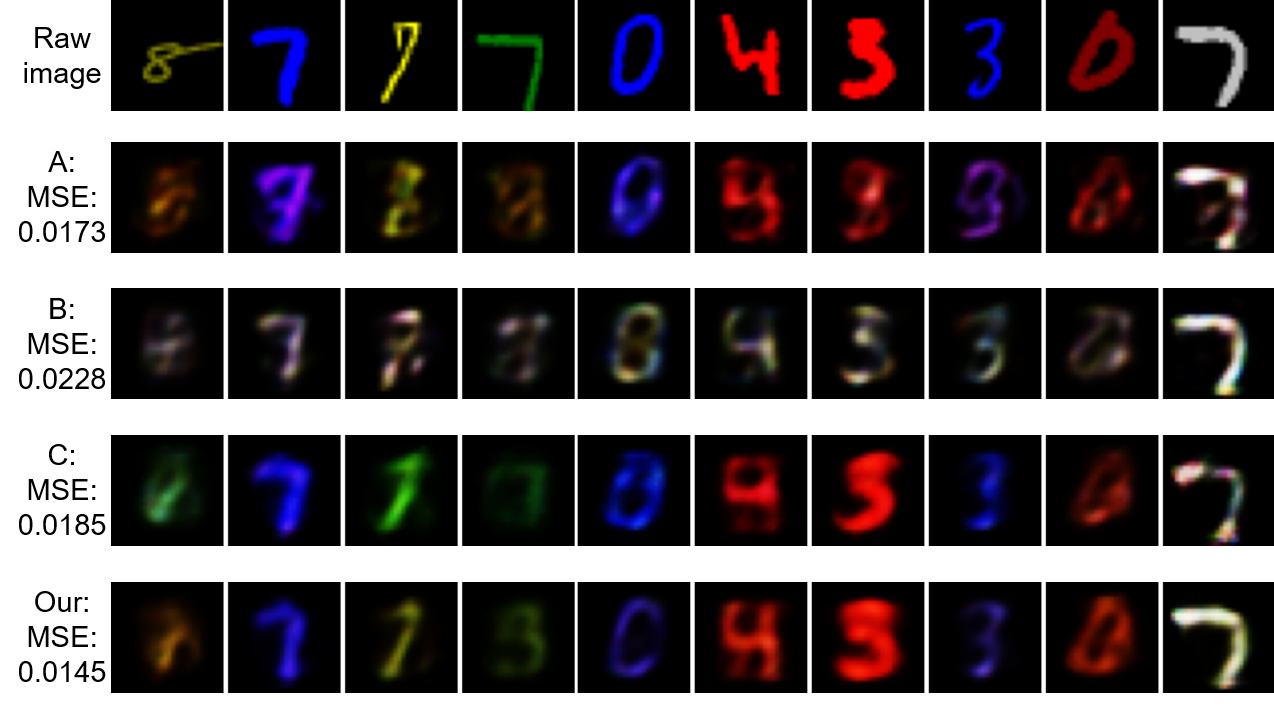}
}
\caption{\label{fig:Reconstructions}Visual comparison between baselines and DIB-PPJSCC on the colored MNIST dataset.}
\end{figure}

\begin{figure}[t]
\centering{
\includegraphics[width=0.9\linewidth]{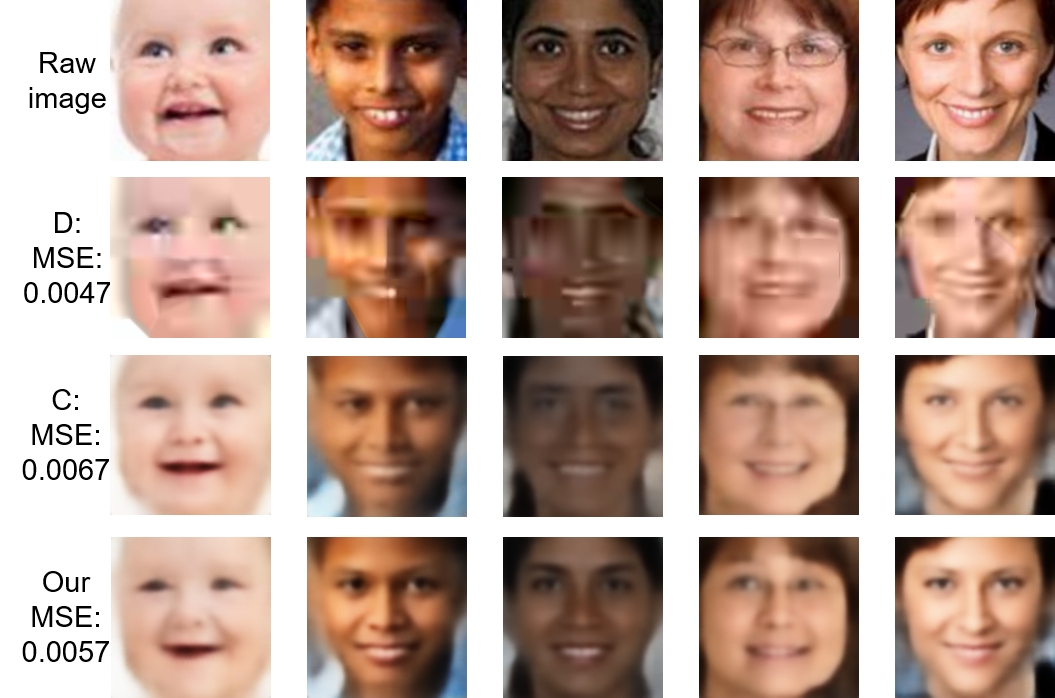}
}
\caption{\label{fig:SSCC_Reconstructions}Visual comparison between separate scheme and the proposed schemes on the UTK face dataset.}
\end{figure}
\subsubsection{Implementation details}
To make a fair comparison, the settings and structure of neural networks used in DIB-PPJSCC are the same as those used in baselines. We choose ${\alpha_1}$ and ${\beta _1}$ from the set $\left\{ {0.5, 1, 1.5, 2, 2.5} \right\} \times {10^5}$, ${p_{{\rm{level}}}}$ from the set $\left\{ {4, 8, 16, 32, 64, 128} \right\}$ that provide the best performance. ${V_{{{\rm{d}}_1}}}$, ${V_{{{\rm{d}}_2}}}$ and ${V_{\rm{p}}}$ are set to be $30$, $50$ and $50$, respectively.

For the colored MNIST dataset, the private encoder is composed of $3$ FC layers with dimensions ${\rm{2352}} \to {\rm{512}} \to 128 \to {M_{\rm{s}}}$, while the public encoder consists of $3$ FC layers with dimensions ${\rm{2352}} \to {\rm{512}} \to 128 \to {M_{\rm{t}}}$. The decoder is constructed with three FC layers of dimensions $M \to 256 \to {\rm{512}} \to {\rm{2352}}$. The optimizers of encoders and the decoder are Adam optimizers with a learning rate of $0.001$ and betas of $\left( {0.9,0.999} \right)$. For the UTK face dataset, the private encoder consists of a convolutional neural network (CNN) with $5$ convolutional layers, followed by an FC layer with dimensions of $512 \times 8 \times 8 \to {M_{\rm{s}}}$. The public encoder consists of a CNN with $5$ convolutional layers followed by an FC layer with dimensions of $512 \times 8 \times 8 \to {M_{\rm{t}}}$. The decoder is constructed with an FC layer of dimension $M \to 512 \times 8 \times 8$ followed by $5$ transposed convolutional layers. The optimizers of encoders and the decoder are Adam optimizers with a learning rate of $0.0002$ and betas of $\left( {0.5,0.999} \right)$. 

For baseline A and C, the JSCC encoder and decoder are jointly trained for ${V_{{{\rm{d}}_2}}}$ epochs. For baseline B, the JSCC encoder and decoder are trained similarly to DIB-PPJSCC following Algorithm \ref{algorithm:DIB}, and then ${{{\boldsymbol{y}}_{\rm{s}}}}$ is fixed to 0, and ${{\rm{D}}_{{{\boldsymbol{\theta }}_{\rm{B}}}}}$ is trained to reduce the reconstruction distortion for $50$ epochs. 

Classifiers consisting of $2$ FC layers with dimensions $M \to 16 \to S$ are trained for $5$ epochs to detect the private information from codewords extracted by the encoder. For baseline D, the MLP classifier is trained using the QAM codewords. The accuracy of these classifiers in classifying private information serves as a metric of defense against eavesdroppers.

\subsection{Performance comparsion}\label{Ablative study}

\begin{figure*}[t!]
\centering 
\setlength{\abovecaptionskip}{0cm}
\setlength{\belowcaptionskip}{-0.5cm}
\subfigbottomskip=7pt 
\subfigcapskip=0pt 
\subfigure[t-SNE on ${{\boldsymbol{y}}_{\rm{t}}}$ with respect to color.]{
\label{fig:TSNE-method1}
\includegraphics[width=.23\linewidth]{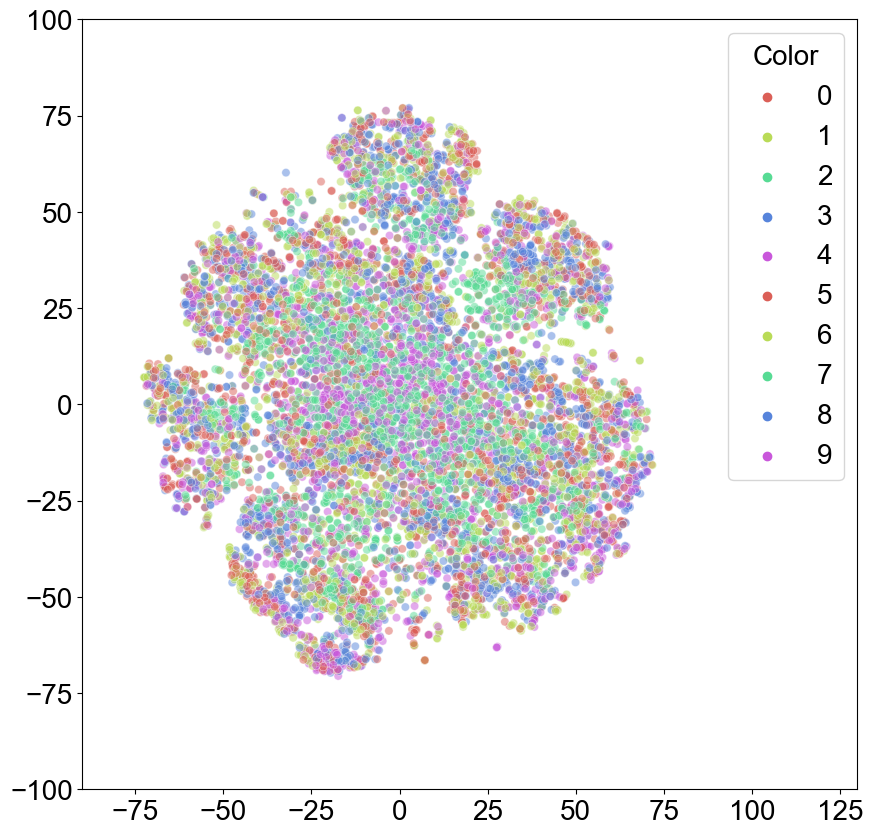}}
\subfigure[t-SNE on ${{\boldsymbol{y}}_{\rm{t}}}$ with respect to digits.]{
\label{fig:TSNE-method2}
\includegraphics[width=.23\linewidth]{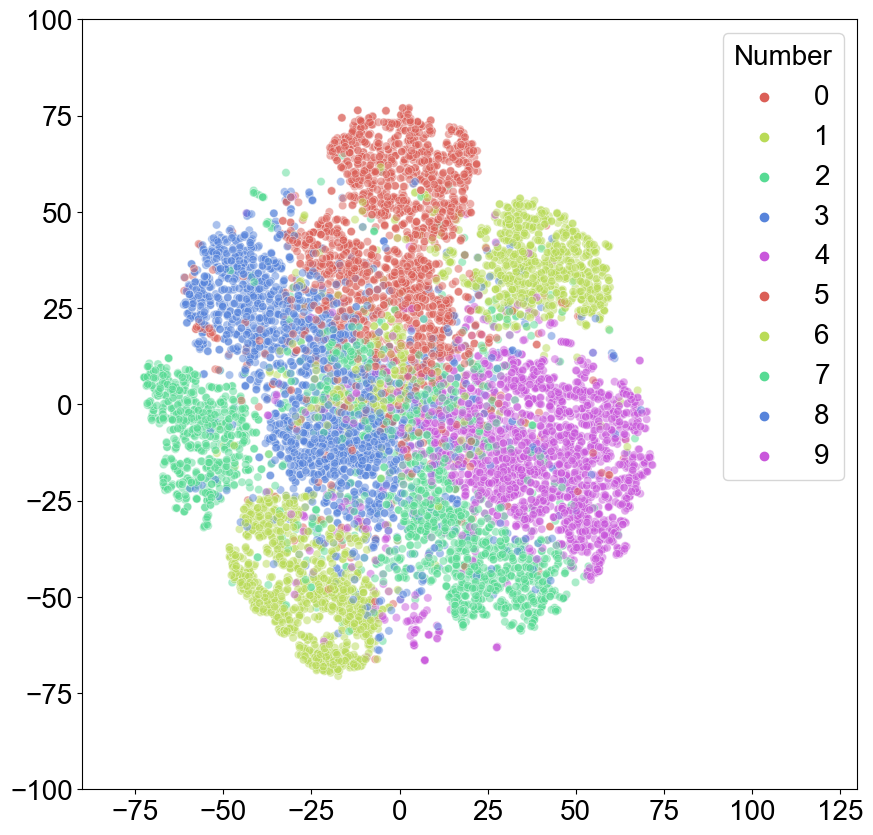}}
\subfigure[t-SNE on ${{\boldsymbol{y}}_{\rm{s}}}$ with respect to color.]{
\label{fig:TSNE-method3}
\includegraphics[width=.23\linewidth]{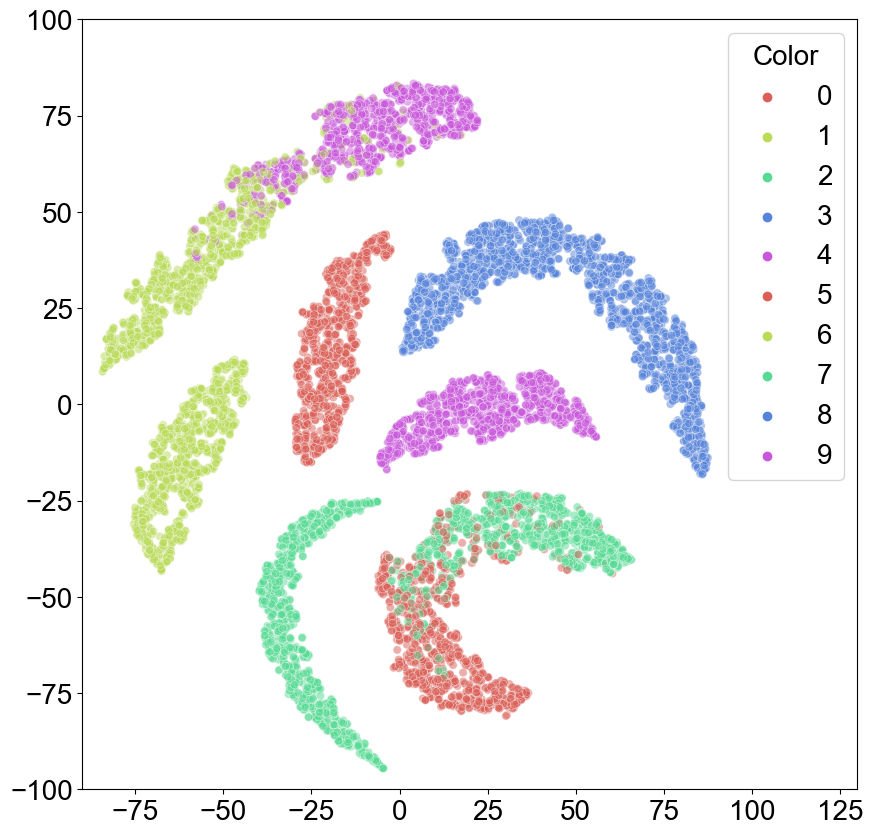}}
\subfigure[t-SNE on ${{\boldsymbol{y}}_{\rm{s}}}$ with respect to digits.]{
\label{fig:TSNE-method4}
\includegraphics[width=.23\linewidth]{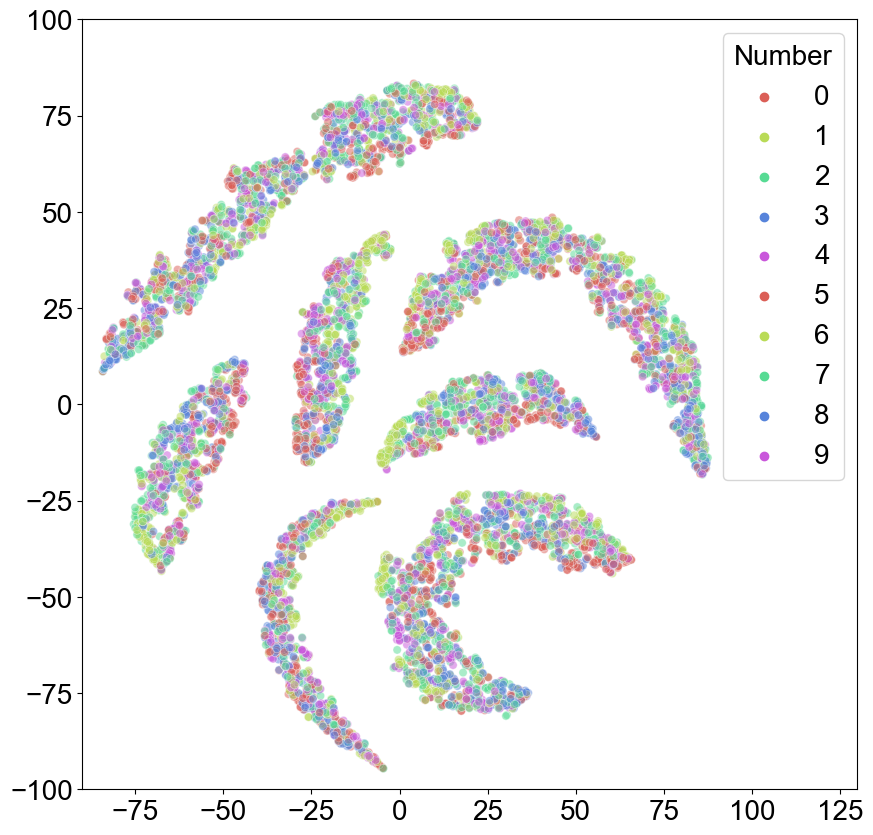}}
\caption{t-SNE visualization of public subcodewords ${{\boldsymbol{y}}_{\rm{t}}}$ and private subcodewords ${{\boldsymbol{y}}_{\rm{s}}}$ extracted by DIB-PPJSCC for test set of the colored MNIST dataset. The neural network is trained when ${\rm{SN}}{{\rm{R}}_{{\rm{AB}}}} = 30\rm{dB}$. Each color represents a different class.}
\label{fig:TSNE-method}
\end{figure*}

Table \ref{tab:Acc} shows the private information classification accuracy, i.e., the eavesdropping accuracy of baselines A, B, C and DIB-PPJSCC under various ${\rm{SN}}{{\rm{R}}_{{\rm{AE}}}}$. ${p_{{\rm{level}}}}$ is fixed to $128$, and ${\rm{Len}}$ used for the colored MNIST and the UTK face datasets are $16$ and $10$, respectively. The reconstruction MSE of all methods is kept close on both datasets, except for baseline B on the colored MNIST dataset. The MSE of baseline B is approximately 0.01 larger than other methods on the colored MNIST dataset due to the failure of baseline B in achieving a lower MSE. From Table \ref{tab:Acc}, we can observe that the eavesdropping accuracy of DIB-PPJSCC is lower than all other methods on the UTK face dataset. On the colored MNIST dataset, the eavesdropping accuracy of DIB-PPJSCC is lower than baseline A and C and close to baseline B  with a smaller MSE. These observations validate the effectiveness of DIB-PPJSCC. From Table \ref{tab:Acc}, we can also observe that the eavesdropping accuracy of baseline B on the colored MNIST dataset is close to random guess (about 0.1 for 10 categories) and exhibits minimal variation when ${\rm{SN}}{{\rm{R}}_{{\rm{AE}}}}$ increases. This is because the color information can be completely separated from other information, and the public codewords on the colored MNIST dataset in baseline B contain no information about color, thus leading to eavesdropping accuracy close to that of a random guess.

\begin{figure*}[t!]
\centering 
\setlength{\abovecaptionskip}{0cm}
\setlength{\belowcaptionskip}{-0.5cm}
\subfigbottomskip=7pt 
\subfigcapskip=0pt 
\subfigure[t-SNE of ${\boldsymbol{y}}_{\rm{s}}$ when ${\rm{SN}}{{\rm{R}}_{{\rm{AE}}}} = 5{\rm{dB}}$.]{
\label{fig:psw1}
\includegraphics[width=.3\linewidth]{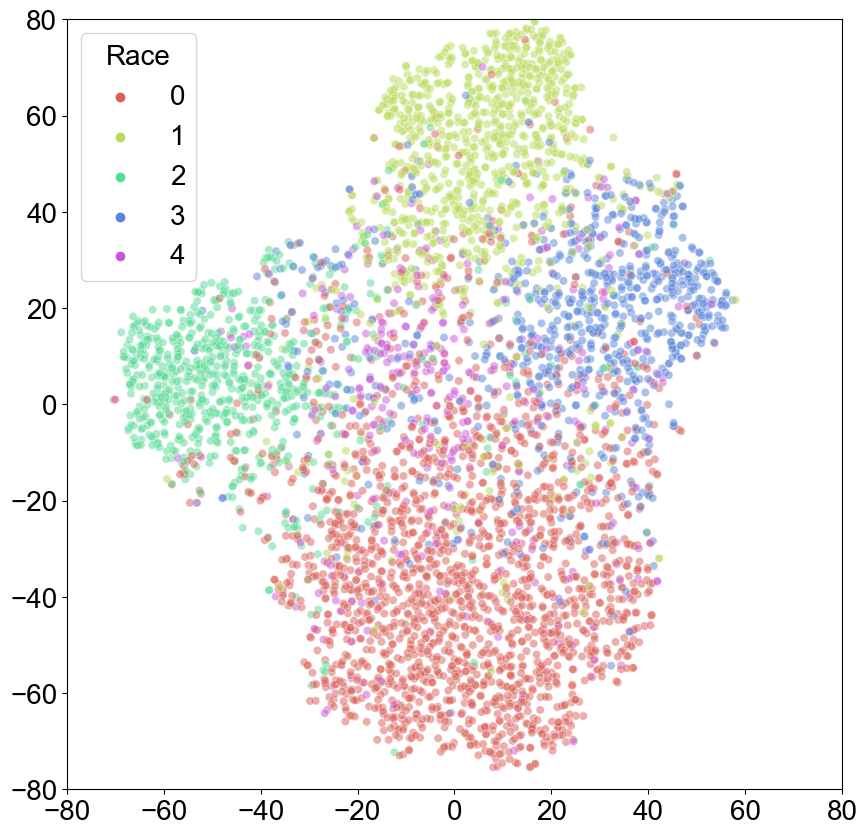}}\hspace{2pt}
\subfigure[t-SNE of ${\boldsymbol{y}}_{\rm{s}}^{\boldsymbol{p}}$ when ${\rm{SN}}{{\rm{R}}_{{\rm{AE}}}} = 5{\rm{dB}}$.]{
\label{fig:psw2}
\includegraphics[width=.3\linewidth]{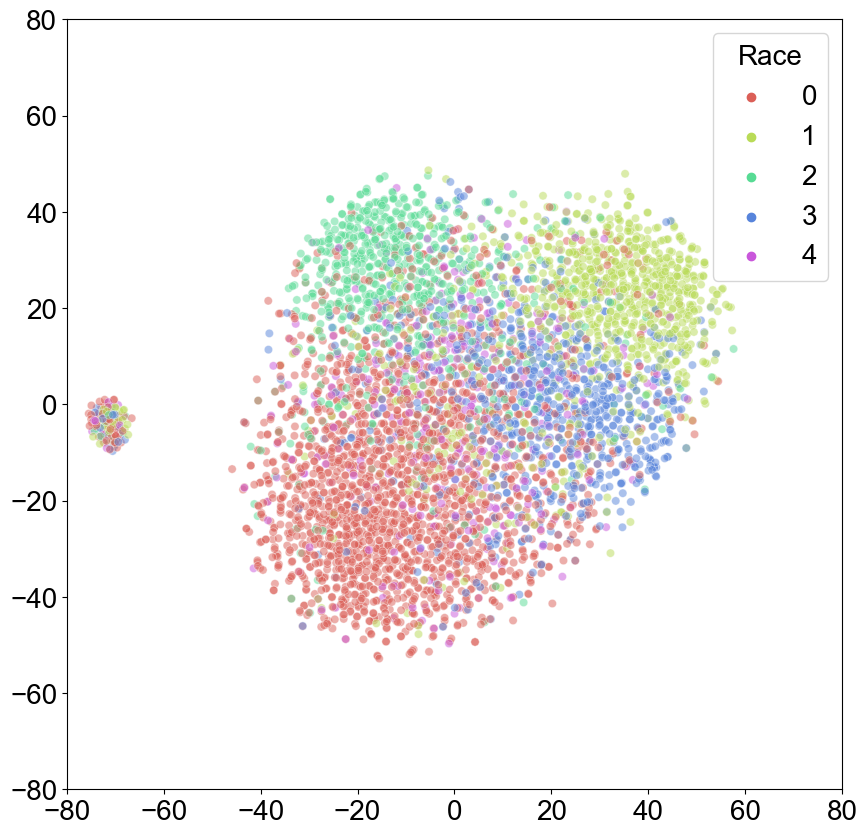}}\hspace{2pt}
\subfigure[t-SNE of ${\boldsymbol{y}}_{\rm{s}}^{\boldsymbol{p}}$ when ${\rm{SN}}{{\rm{R}}_{{\rm{AE}}}} = 15{\rm{dB}}$.]{
\label{fig:psw3}
\includegraphics[width=.3\linewidth]{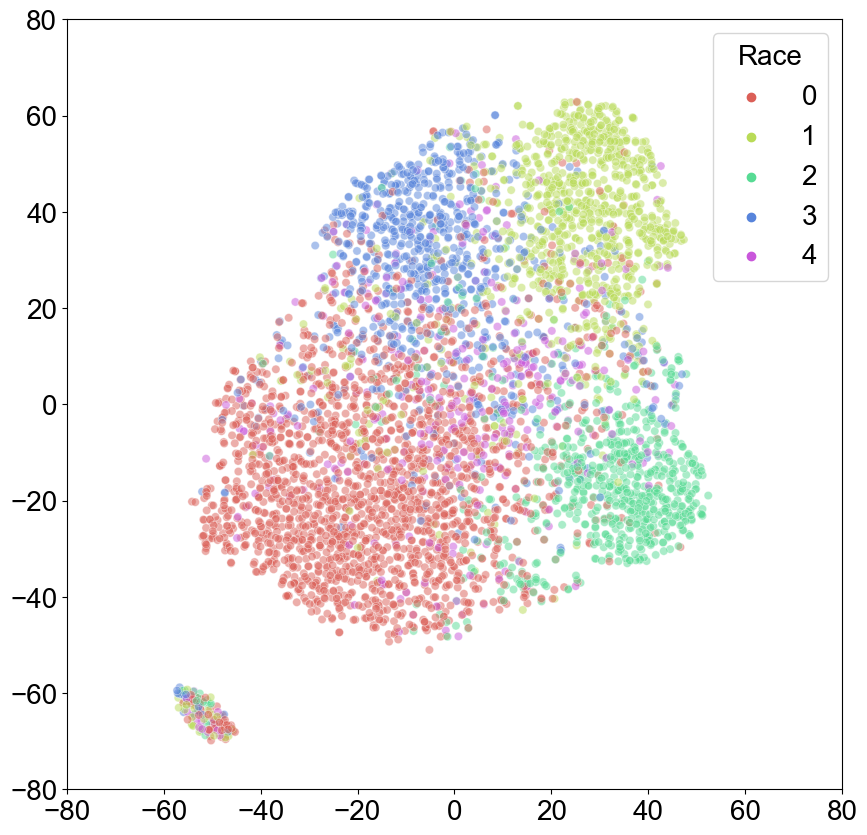}}
\caption{t-SNE visualization of the private subcodeword ${\boldsymbol{y}}_{\rm{s}}$ and the protected private subcodeword ${\boldsymbol{y}}_{\rm{s}}^{\boldsymbol{p}}$ extracted by DIB-PPJSCC for test set of the UTK dataset. The neural network is trained when ${\rm{SN}}{{\rm{R}}_{{\rm{AB}}}} = 30\rm{dB}, {\rm{SN}}{{\rm{R}}_{{\rm{AE}}}} = 5\rm{dB}, 15\rm{dB}$. Each color represents a different class of race.}
\label{fig:psw}
\end{figure*}

Figure \ref{fig:Reconstructions} shows the visual reconstructions of baselines A, B, C and DIB-PPJSCC on the colored MNIST dataset when ${\rm{SN}}{{\rm{R}}_{{\rm{AB}}}} = 30\rm{dB}$ and ${\rm{SN}}{{\rm{R}}_{{\rm{AE}}}} = 5\rm{dB}$. From Fig. \ref{fig:Reconstructions}, we can observe that the reconstruction error and the visual performance of DIB-PPJSCC are both superior to all other baselines. For example, in the third column of the fifth row of Fig. \ref{fig:Reconstructions}, the color can be clearly identified as yellow, while in the third column of other rows, the color may be identified as green or white. Hence, DIB-PPJSCC can protect private information and recover images more precisely. From Fig. \ref{fig:Reconstructions}, we can also observe that the reconstruction error of baseline B is much larger than those of other methods even though baseline B has the lowest eavesdropping accuracy. This is because baseline B discards all color-relevant information in private codewords, and the color of all recovered images is similar, shown as in the third row of Fig. \ref{fig:Reconstructions}. We can also observe that although losing color-relevant information, baseline B can keep consistent with the raw image in other information such as the value of handwritten digits. For example, in the second column of the third row of Fig. \ref{fig:Reconstructions}, we can identify the digit as 7 although the color is different from that in the corresponding raw image. This makes baseline B valuable when the security requirements are strict since baseline B is able to thoroughly discard the private information without losing public information in the recovered images. 

Figure \ref{fig:SSCC_Reconstructions} shows the visual reconstructions of baselines D, C and DIB-PPJSCC on the UTK face dataset when ${\rm{SN}}{{\rm{R}}_{{\rm{AB}}}} = 30{\rm{dB}}$ and ${\rm{SN}}{{\rm{R}}_{{\rm{AE}}}} = 5\rm{dB}$.  Since we choose the password uniformly from ${\left[ {1, \ldots, {p_{{\rm{level}}}}} \right]^{{\rm{Len}}}}$, the information amount of the password used by DIB-PPJSCC is 
\begin{equation}
\label{experiment-SSCC-1}
{H_{\boldsymbol{p}}} = {\rm{Len}} \times \log _2^{{p_{{\rm{level}}}}} \left( {{\rm{bits}}} \right).
\end{equation} 
${p_{{\rm{level}}}}$ and ${\rm{Len}}$ are set to be $256$ and $16$ to keep the consistency between the passwords used by baseline D and DIB-PPJSCC (both 128 bits). According to Shannon’s separation theorem \cite{shannon1948mathematical}, the necessary and sufficient condition for reliable communication over a channel with capacity $C$ is that the transmitted rate $R$ is smaller than its upper bound ${R_{\max }}$, i.e.,
\begin{equation}
\label{experiment-SSCC-2}
R \le {R_{\max }} = \frac{M}{N}C.
\end{equation} 
For AWGN channel, there is $C = \log _2^{\left( {1 + {\rm{SNR}}} \right)}$. Then, we calculate the maximum achievable rate, $R_{\max }$ as the compression level utilized by the BPG encoder. The eavesdropping accuracy of the baseline D is around $0.35$ under different ${\rm{SN}}{{\rm{R}}_{{\rm{AE}}}}$ due to the randomness in the QAM codewords generated through AES encryption scheme. The eavesdropping accuracy of DIB-PPJSCC ranges from $0.2166$ to $0.4219$ and is higher than that of the separate scheme only when ${\rm{SN}}{{\rm{R}}_{{\rm{AE}}}} = 10, 15\rm{dB}$. From Fig. \ref{fig:SSCC_Reconstructions}, we can observe that even though the reconstruction MSE of baseline D is slightly lower than baseline C and DIB-PPJSCC, the key facial information such as eye and nose positions is damaged severely. This is because the BPG breaks down the adjacent pixels of the raw image into multiple blocks for compression without considering the semantic relationship between pixels, thus potentially damaging the key information. In addition, even though the eavesdropping accuracy of DIB-PPJSCC is slightly higher than that of baseline D when ${\rm{SN}}{{\rm{R}}_{{\rm{AE}}}} = 10\rm{dB}, 15\rm{dB}$, baseline D exhibits the cliff effect when ${\rm{SN}}{{\rm{R}}_{{\rm{AB}}}}$ decreases due to the thresholding effects of the channel code\cite{bourtsoulatze2019deep}. Consequently, the recovered images cannot be opened due to damage to the BPG header. Hence, the DIB-PPJSCC has better reconstruction and robustness performance than the separate scheme.

\begin{table*}[t!]
\setlength\tabcolsep{5pt}
\renewcommand\arraystretch{1.1}
\centering
\caption{\label{tab:roub1} Eavesdropping accuracy under different ${\rm{SN}}{{\rm{R}}_{{\rm{AB}}}}$ and ${\rm{SN}}{{\rm{R}}_{{\rm{AE}}}}$.}
\begin{tabular}{c|c|c|ccccccc}
\toprule
\multirow{2}*{\textbf{Datasets}} & \multirow{2}*{\textbf{Methods}} & \multirow{2}*{\textbf{Train ${\textbf{SN}}{{\textbf{R}}_{{\textbf{AE}}}}$}} & \multicolumn{7}{c}{\textbf{Acc under different test ${\textbf{SN}}{{\textbf{R}}_{{\textbf{AE}}}}$} } \\
~ & ~ &  & -15dB & -10dB & -5dB & 0dB & 5dB & 10dB & 15dB\\
\midrule
\multirow{6}*{Colored MNIST} & \multirow{3}*{Adversarial JSCC} & -15dB & 0.125 & 0.2109 & 0.2578 & 0.3828 & 0.5313 & 0.5938 & 0.6797 \\
~ & ~ & 5dB & 0.1406 & 0.2344 & 0.2422 & 0.3125 & 0.3125 & 0.375 & 0.4922\\
~ & ~ & 15dB & 0.125 & 0.1484 & 0.2031 & 0.2813 & 0.3281 & 0.3203 & 0.3647 \\
\cmidrule{2-10}
~ & \multirow{3}*{DIB-PPJSCC} & -15dB & 0.0938 & 0.1485 & 0.1485 & 0.1641 & 0.2656 & 0.3906 & 0.5938\\
~ & ~ & 5dB & 0.1172 & 0.125 & 0.1641 & 0.1563 & 0.1328 & 0.3203 & 0.4688\\
~ & ~ & 15dB & 0.1172 & 0.1016 & 0.1328 & 0.1328 & 0.1406 & 0.1875 & 0.1953\\
\midrule
\multirow{6}*{UTK face} & \multirow{3}*{Adversarial JSCC} & -15dB & 0.2344 & 0.4375 & 0.4688 & 0.5 & 0.4844 & 0.5 & 0.6094\\
~ & ~ & 5dB & 0.3438 & 0.4375 & 0.4375 & 0.5469 & 0.4063 & 0.4844 & 0.4844\\
~ & ~ & 15dB & 0.3906 & 0.4531 & 0.4531 & 0.5 & 0.5 & 0.6094 & 0.4844\\
\cmidrule{2-10}
~ & \multirow{3}*{DIB-PPJSCC} & -15dB & 0.2188 & 0.3437 & 0.3593 & 0.4688 & 0.4531 & 0.5 & 0.5312\\
~ & ~ & 5dB & 0.2967 & 0.2813 & 0.3438 & 0.4531 & 0.375 & 0.4375 & 0.4844\\
~ & ~ & 15dB & 0.3218 & 0.3594 & 0.4219 & 0.4688 & 0.3906 & 0.4844 & 0.4063\\
\bottomrule
\end{tabular}
\end{table*}

\subsection{Qualitative analysis}\label{Qualitative Analysis}
Figure \ref{fig:TSNE-method} shows the 2-dimensional projections of the public subcodewords ${{\boldsymbol{y}}_{\rm{t}}}$ and private subcodewords ${{\boldsymbol{y}}_{\rm{s}}}$ extracted by DIB-PPJSCC for test set of the colored MNIST dataset. In particular, we utilize t-Distributed Stochastic Neighbor Embedding (t-SNE)\cite{van2008visualizing} to project the codewords into a 2-dimensional space. We also show the labels of each image with regard to the private information, i.e., color, and the public information, i.e., digits, to make it easier to investigate the clusters. From Fig. \ref{fig:TSNE-method1} and \ref{fig:TSNE-method2}, we can observe that the t-SNE projections of ${{\boldsymbol{y}}_{\rm{t}}}$ with different colors exhibit significant overlap, while the t-SNE projections of ${{\boldsymbol{y}}_{\rm{t}}}$ with different digits are separated well. This indicates that ${{\boldsymbol{y}}_{\rm{t}}}$ contains abundant digit-related information and is agnostic to the private information. From Fig. \ref{fig:TSNE-method3} and \ref{fig:TSNE-method4}, we can also observe that the t-SNE projections of ${{\boldsymbol{y}}_{\rm{s}}}$ shows distributions totally different to those of ${{\boldsymbol{y}}_{\rm{t}}}$. The t-SNE projections of ${{\boldsymbol{y}}_{\rm{s}}}$ with different colors are distinctly separated, while the t-SNE projections of ${{\boldsymbol{y}}_{\rm{s}}}$ with different digits are mixed together. This suggests that ${{\boldsymbol{y}}_{\rm{s}}}$ contains abundant color-related information while almost no digit-related information. This is because the DIB algorithm shown in Algorithm \ref{algorithm:DIB} preserves as much private information, i.e., the information related to the color in ${{\boldsymbol{y}}_{\rm{s}}}$ as possible, and removes the private information in ${{\boldsymbol{y}}_{\rm{t}}}$ at the same time. In addition, to guarantee the reconstruction quality, the DIB algorithm preserves the public information, i.e., the information related to the digit, that will not cause privacy leakage in ${{\boldsymbol{y}}_{\rm{t}}}$ instead of ${{\boldsymbol{y}}_{\rm{s}}}$, as ${{\boldsymbol{y}}_{\rm{t}}}$ is directly transmitted to the legitimate receiver Bob while ${{\boldsymbol{y}}_{\rm{s}}}$ is altered to be protected using Algorithm \ref{algorithm: PP}. Hence, the DIB algorithm is able to effectively disentangle the private and public information and preserve them in the proper codewords to achieve privacy protection.

Figure \ref{fig:psw} shows the 2-dimensional t-SNE projections of the private subcodewords ${{\boldsymbol{y}}_{\rm{s}}}$ and the protected private subcodewords ${\boldsymbol{y}}_{\rm{s}}^{\boldsymbol{p}}$ extracted by DIB-PPJSCC for test set of the UTK face dataset. The labels of the private information, i.e., race are also shown to investigate the clusters. From Fig. \ref{fig:psw1}, we can observe that the t-SNE projections of ${{\boldsymbol{y}}_{\rm{s}}}$ are clustered based on the class of race, which indicates that ${{\boldsymbol{y}}_{\rm{s}}}$ contains abundant race-related information by applying DIB algorithm. From Fig. \ref{fig:psw2} and Fig. \ref{fig:psw3}, we can observe that the distance between the t-SNE projections of ${\boldsymbol{y}}_{\rm{s}}^{\boldsymbol{p}}$ with different races are smaller than those of ${{\boldsymbol{y}}_{\rm{s}}}$. This is because the PP algorithm reduces the mutual information between ${\boldsymbol{y}}_{\rm{s}}^{\boldsymbol{p}}$ and the label of race. Consequently, ${\boldsymbol{y}}_{\rm{s}}^{\boldsymbol{p}}$ contains less race-related information than ${\boldsymbol{y}}_{\rm{s}}$, and the difference between ${\boldsymbol{y}}_{\rm{s}}^{\boldsymbol{p}}$ with different races are also reduced. Moreover, we can also observe that compared with Fig. \ref{fig:psw1}, some of the t-SNE projections of ${\boldsymbol{y}}_{\rm{s}}^{\boldsymbol{p}}$ are totally mixed together in Fig. \ref{fig:psw2} (the left cluster) and Fig. \ref{fig:psw3} (the bottom cluster) regardless of races. This is because the PP algorithm reduces the KL divergence between the uniform distribution and ${\boldsymbol{y}}_{\rm{s}}^{\boldsymbol{p}}$, and thus increases the randomness of ${\boldsymbol{y}}_{\rm{s}}^{\boldsymbol{p}}$. The left cluster of Fig. \ref{fig:psw2} and the bottom cluster of Fig. \ref{fig:psw3} represent ${\boldsymbol{y}}_{\rm{s}}^{\boldsymbol{p}}$ that are randomly distributed with respect to the race after the application of the PP algorithm. These ${\boldsymbol{y}}_{\rm{s}}^{\boldsymbol{p}}$ with different races are projected to the same space, which demonstrates the effectiveness of the PP algorithm. Therefore, the PP algorithm can reduce privacy leakage as well as increase the randomness of the codewords to be transmitted through the channel.

\begin{figure*}[t!]
\centering
\begin{minipage}[t]{1\linewidth}
\centering
\includegraphics[width=0.75\linewidth]{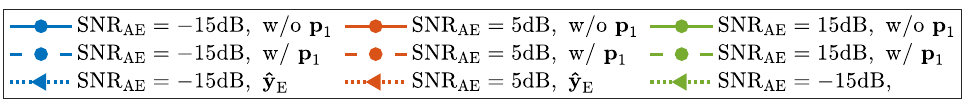}
\end{minipage}
\subfigure[${\rm{Len}} = 4$]{
\label{fig:len_level1}
\includegraphics[width=0.43\linewidth]{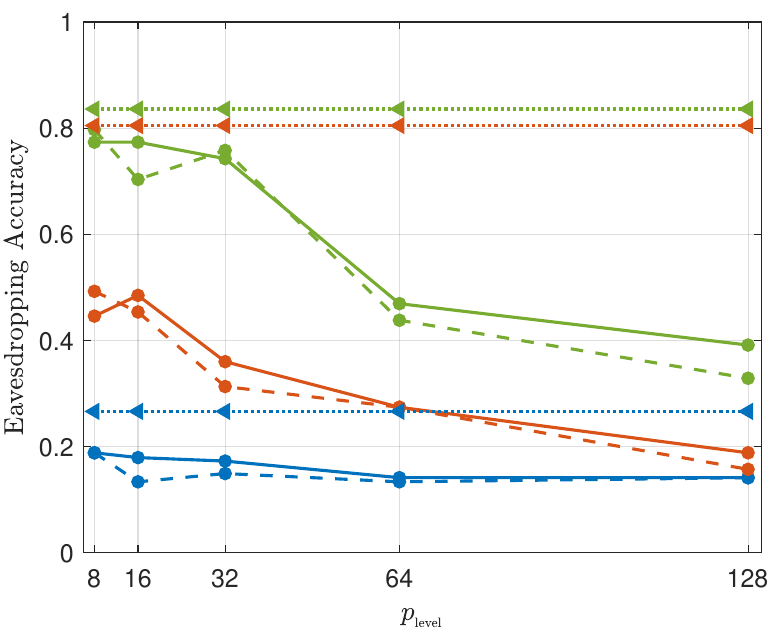}}
\subfigure[${\rm{Len}} = 8$]{
\label{fig:len_level2}
\includegraphics[width=0.43\linewidth]{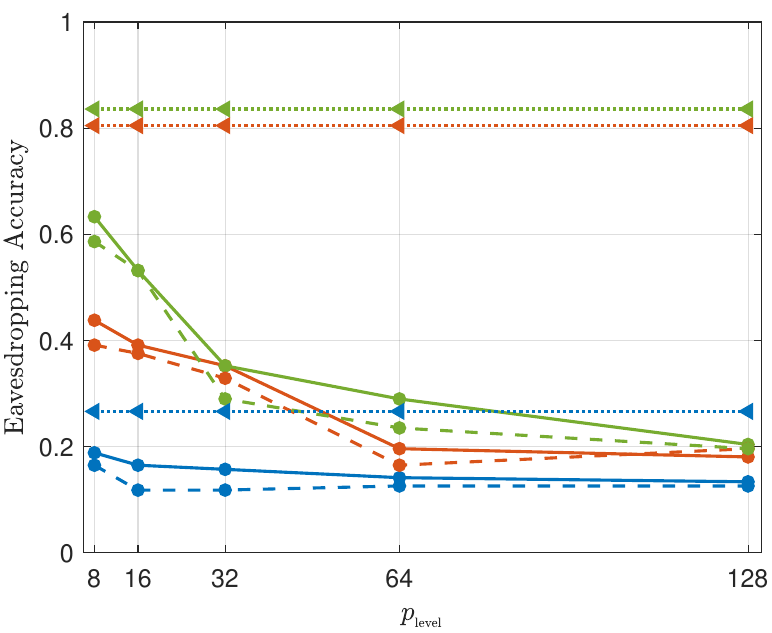}}
\subfigure[${\rm{Len}} = 12$]{
\label{fig:len_level3}
\includegraphics[width=0.43\linewidth]{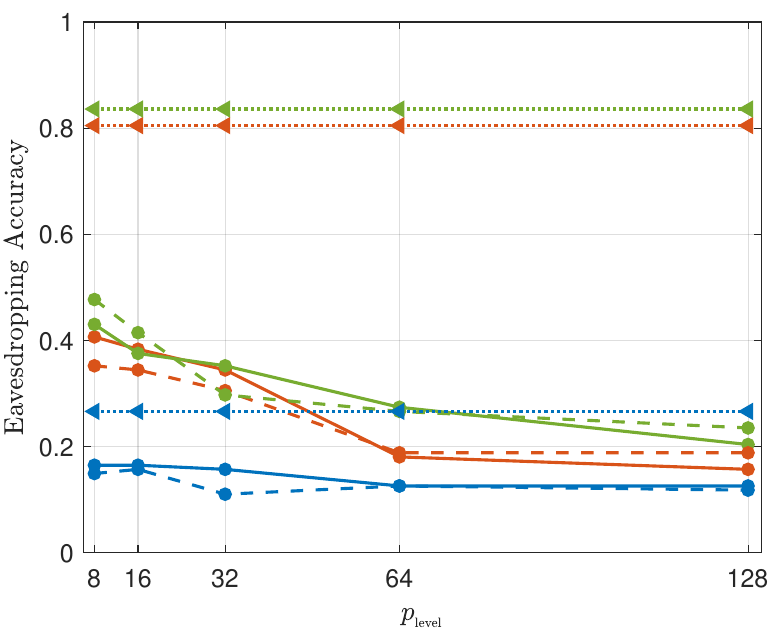}}
\subfigure[${\rm{Len}} = 16$]{
\label{fig:len_level3}
\includegraphics[width=0.43\linewidth]{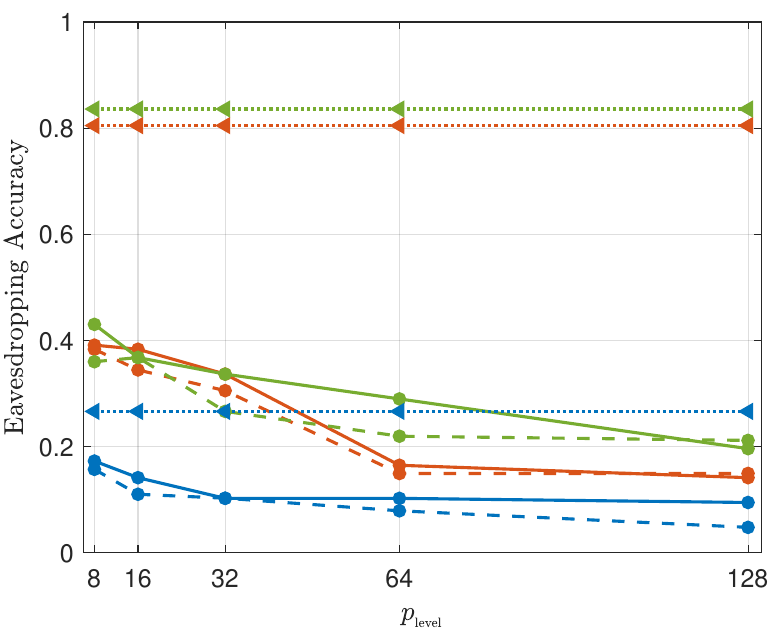}}
\caption{The eavesdropping accuracy on the colored MNIST dataset under various ${\rm{Len}}$ and ${p_{{\rm{level}}}}$. The neural networks are trained when ${\rm{SN}}{{\rm{R}}_{{\rm{AB}}}} = 30{\rm{dB, SN}}{{\rm{R}}_{{\rm{AE}}}} =  - 15{\rm{dB}}, 5{\rm{dB}}, 15{\rm{dB}}$. The solid lines, the dashed lines and the dotted lines represent the eavesdropping accuracy of ${{{\boldsymbol{\hat y}}}_{{\rm{t,E}}}}$ (expressed as w/o ${{\boldsymbol{p}}_{\rm{1}}}$), ${\boldsymbol{\bar y}}_{\rm{E}}^{{\boldsymbol{p}},{{\boldsymbol{p}}_1}}$ (expressed as w/ ${{\boldsymbol{p}}_{\rm{1}}}$) and ${{{\boldsymbol{\hat y}}}_{\rm{E}}}$ (expressed as ${{{\boldsymbol{\hat y}}}_{\rm{E}}}$ ), respectively.}
\label{fig:len_level}
\vspace{-3ex}
\end{figure*}

\subsection{Robustness and complexity}\label{Robust}
Table \ref{tab:roub1} shows the eavesdropping accuracy of DIB-PPJSCC and baseline C on the colored MNIST dataset that is trained when ${\rm{SN}}{{\rm{R}}_{{\rm{AB}}}} = 30{\rm{dB, SN}}{{\rm{R}}_{{\rm{AE}}}} =  - 15{\rm{dB}}, 5{\rm{dB}}, 15{\rm{dB}}$ and tested under different ${\rm{SN}}{{\rm{R}}_{{\rm{AE}}}}$. From Table \ref{tab:roub1}, we can observe that the eavesdropping accuracy of DIB-PPJSCC is always lower than that of baseline C. This implies that DIB-PPJSCC has better robustness than baseline C when there is an estimated error on ${\rm{SN}}{{\rm{R}}_{{\rm{AE}}}}$. From Table \ref{tab:roub1}, we can also observe that increasing the train ${\rm{SN}}{{\rm{R}}_{{\rm{AE}}}}$ can reduce the privacy leakage caused by the mismatching between train and test ${\rm{SN}}{{\rm{R}}_{{\rm{AE}}}}$. For instance, when the test ${\rm{SN}}{{\rm{R}}_{{\rm{AE}}}} = 15{\rm{dB}}$, the 
eavesdropping accuracy when the train ${\rm{SN}}{{\rm{R}}_{{\rm{AE}}}} = -15{\rm{dB}}$, increases up to 0.5938 and 0.5312 for the colored MNIST and UTK face datasets, respectively. However, the eavesdropping accuracy is 0.4688 and 0.4844 when the train ${\rm{SN}}{{\rm{R}}_{{\rm{AE}}}} = 5{\rm{dB}}$, which is lower than 0.5938 and 0.5312. Moreover, when the train ${\rm{SN}}{{\rm{R}}_{{\rm{AE}}}} = 15{\rm{dB}}$ and the test ${\rm{SN}}{{\rm{R}}_{{\rm{AE}}}}$ varies, the eavesdropping accuracy slightly fluctuates around $0.15$ on the colored MNIST dataset, and $0.4$ on the UTK face dataset, which is close to the random guess probability. This is because when the train ${\rm{SN}}{{\rm{R}}_{{\rm{AE}}}}$ is larger, the randomness of ${{\boldsymbol{y}}_{\rm{s}}}$ introduced by the noise becomes weaker. To prevent privacy leakage, the PP algorithm needs to impose a stronger security constraint on ${{\boldsymbol{y}}_{\rm{s}}}$. Hence, the defense capability against eavesdropping is better than that when the train ${\rm{SN}}{{\rm{R}}_{{\rm{AE}}}}$ is smaller. This indicates that DIB-PPJSCC has better robustness to the change of test ${\rm{SN}}{{\rm{R}}_{{\rm{AE}}}}$ when the train ${\rm{SN}}{{\rm{R}}_{{\rm{AE}}}}$ is larger.

Figure \ref{fig:len_level} shows the eavesdropping accuracy of ${{{\boldsymbol{\hat y}}}_{\rm{E}}}$ as well as the eavesdropping accuracy of ${{{\boldsymbol{\hat y}}}_{{\rm{t,E}}}}$ and ${\boldsymbol{\bar y}}_{{\rm{s,E}}}^{{\boldsymbol{p}},{{\boldsymbol{p}}_1}}$ on the colored MNIST dataset under different password lengths ${{\rm{Len}}}$ and password levels ${p_{{\rm{level}}}}$ when ${\rm{SN}}{{\rm{R}}_{{\rm{AB}}}} = 30{\rm{dB}}$ and ${\rm{SN}}{{\rm{R}}_{{\rm{AE}}}} =  - 15,5,15{\rm{dB}}$.  From Fig. \ref{fig:len_level}, we can observe that when ${{\rm{Len}}}$ and ${p_{{\rm{level}}}}$ become larger, the eavesdropping accuracy becomes smaller. This is because the larger the ${{\rm{Len}}}$ and ${p_{{\rm{level}}}}$ are, the randomness of ${{\boldsymbol{y}}_{\rm{s}}}$ introduced by the PP algorithm is stronger, and thus decreasing the eavesdropping accuracy more. Moreover, the eavesdropping accuracy of ${\boldsymbol{\bar y}}_{{\rm{s,E}}}^{{\boldsymbol{p}},{{\boldsymbol{p}}_1}}$ which is recovered by the eavesdropper using the wrong password ${{{\boldsymbol{p}}_1}}$, is always close to the eavesdropping accuracy of ${{{\boldsymbol{\hat y}}}_{{\rm{t,E}}}}$ under all ${{\rm{Len}}}$ and ${p_{{\rm{level}}}}$. This demonstrates the effectiveness of the PP algorithm in defending against the strong eavesdropper that has access to ${\rm{T}}_\phi ^{ - 1}$. From Fig. \ref{fig:len_level}, we can also observe that the eavesdropping accuracy will be significantly reduced as long as using the PP algorithm to protect private information no matter what ${{\rm{Len}}}$ and ${p_{{\rm{level}}}}$ are. For instance, when ${\rm{Len = 4}}$, ${p_{{\rm{level}}}} = 8$ and ${\rm{SN}}{{\rm{R}}_{{\rm{AE}}}} = 15{\rm{dB}}$ where the eavesdropper is strong and the password is poorly random, the eavesdropping accuracy of DIB-PPJSCC is still lower than that of ${{{\boldsymbol{\hat y}}}_{\rm{E}}}$. This implies that the PP algorithm is robust to ${{\rm{Len}}}$ and ${p_{{\rm{level}}}}$, and we can apply different ${{\rm{Len}}}$ and ${p_{{\rm{level}}}}$ to achieve various security levels.

\begin{figure*}[t!]
\centering 
\setlength{\abovecaptionskip}{0cm}
\setlength{\belowcaptionskip}{-0.5cm}
\subfigbottomskip=7pt 
\subfigcapskip=0pt 
\subfigure[The colored MNIST dataset.]{
\label{fig:snrAB1}
\includegraphics[width=.4\linewidth]{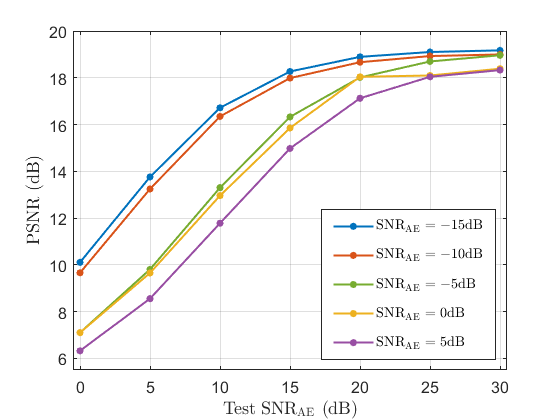}}
\subfigure[The UTK face dataset.]{
\label{fig:snrAB2}
\includegraphics[width=.4\linewidth]{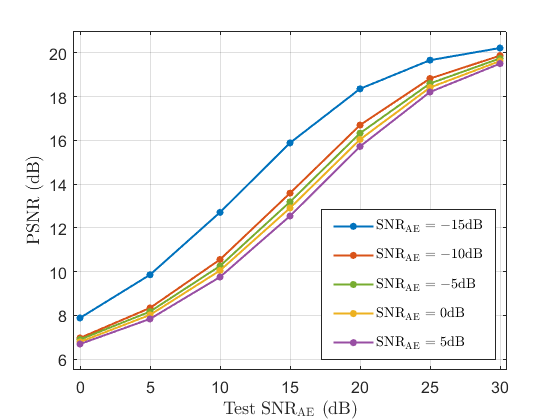}}
\caption{The PSNR of the images recovered by DIB-PPJSCC for the test set of the colored MNIST and the UTK face dataset. The neural network is trained when ${\rm{SN}}{{\rm{R}}_{{\rm{AB}}}} = 30\rm{dB}, {\rm{SN}}{{\rm{R}}_{{\rm{AE}}}} = -5\rm{dB}, 0\rm{dB}, 5\rm{dB}, 10\rm{dB}, 15\rm{dB}$ and is tested under various ${\rm{SN}}{{\rm{R}}_{{\rm{AB}}}}$.}
\label{fig:snrAB}
\end{figure*} 

Figure \ref{fig:snrAB} shows the peak signal-to-noise ratio (PSNR) of DIB-PPJSCC under various test ${\rm{SN}}{{\rm{R}}_{{\rm{AB}}}}$ on test sets of the colored MNIST and the UTK face datasets. PSNR is
\begin{equation}
\label{PSNR}
{\rm{PSNR}} = 10{\log _{10}}\left( {\frac{{{{\left( {{{\rm{2}}^n} - 1} \right)}^2}}}{{{\rm{mse}}\left( {{\boldsymbol{x}},{\boldsymbol{\hat x}}} \right)}}} \right),
\end{equation}
where $n$ is the number of bits that each image pixel uses, ${\rm{mse}}\left( {{\boldsymbol{x}},{\boldsymbol{\hat x}}} \right)$ is MSE between ${\boldsymbol{x}}$ and ${{\boldsymbol{\hat x}}}$. In particular, ${\rm{SN}}{{\rm{R}}_{{\rm{AB}}}}$ is fixed as $30{\rm{dB}}$ when training. The neural networks are trained with identical hyperparameters under various ${\rm{SN}}{{\rm{R}}_{{\rm{AE}}}}$ and tested under various ${\rm{SN}}{{\rm{R}}_{{\rm{AB}}}}$. From Fig. \ref{fig:snrAB}, we can observe that the PSNR decreases when the train ${\rm{SN}}{{\rm{R}}_{{\rm{AE}}}}$ increases under all test ${\rm{SN}}{{\rm{R}}_{{\rm{AB}}}}$. This is because an increase in the train ${\rm{SN}}{{\rm{R}}_{{\rm{AE}}}}$ means stronger eavesdroppers. In order to defend against eavesdropping, the codewords extracted by DIB-PPJSCC will pay more attention to privacy protection, thus reducing the reconstruction quality. From Fig. \ref{fig:snrAB}, we can also observe that when the test ${\rm{SN}}{{\rm{R}}_{{\rm{AB}}}}$ drops below the train ${\rm{SN}}{{\rm{R}}_{{\rm{AB}}}}$, the performance does not saturate immediately, and the reconstruction quality exhibits a graceful degradation. This is because DIB-PPJSCC uses the channel conditions in the loss function and enables the learned codewords to resist channel interference. Hence, the codewords extracted by DIB-PPJSCC are robust to different test ${\rm{SN}}{{\rm{R}}_{{\rm{AB}}}}$.

We also test the time and calculate the computational complexity of baselines and DIB-PPJSCC. The average run time achieved by DIB-PPJSCC on the CPU is 300ms per image. The neural networks used by DIB-PPJSCC consist of convolutions/deconvolutions and FC layers. The computational cost of a single convolutional layer is $H \times W \times K \times K \times {C_i} \times {C_o}$, where $K$ is the filter size, ${C_o}$ is the number of output channels, ${C_i}$ is the number of input channels and $H \times W$ is the size of the feature map. The computational cost of an FC layer is $\left( {2I - 1} \right)O$, where $I$ is the input vector dimension and $O$ is the output vector dimension. Only the width and height of the feature map and the vector dimension depend on the image dimensions, and all other factors are constant and independent of the image size. Consequently, the computational complexity of DIB-PPJSCC is linear with respect to the pixel count of the input image. Regarding the DL-based baselines (A, B and C), the time and computational complexity are similar since the neural networks used by these baselines are similar to those used by DIB-PPJSCC. For the separate scheme, the encoding and decoding time is 337ms per image, which is larger than DIB-PPJSCC due to the iterative channel decoding process required by LDPC to achieve optimal error correction capabilities. Furthermore, it is worth noting that the separate schemes will need more time for optimal iterative channel decoding when ${\rm{SN}}{{\rm{R}}_{{\rm{AB}}}}$ becomes smaller\cite{hagenauer1996iterative,vikalo2004iterative}. The above results prove that DIB-PPJSCC is competitive with the baselines not only in terms of reconstruction quality and privacy protection but also in terms of complexity.

\section{Conclusion}
In this work, we have proposed a DIB-PPJSCC scheme for privacy-protective image transmission, which can protect private information as well as recover it at the legitimate receiver. Specifically, we first derived a mathematically tractable form of the IB objective for disentangling private information and public information. Then, to appropriately protect private information, we further proposed a password-based privacy-protective algorithm that can prevent privacy leakage and guarantee privacy recovery simultaneously. Experimental results have shown that DIB-PPJSCC always achieved a better balance between privacy protection and reconstruction quality than other baselines, which demonstrates the effectiveness of the DIB objective and the password-based privacy-protective algorithm. In addition, the images recovered by DIB-PPJSCC had better visual performance and preserved more critical information than other baselines. The eavesdropping accuracy of DIB-PPJSCC gradually decreases when the level and length of the passwords increase, which implies that the PP algorithm is robust and is able to provide different levels of privacy protection. When the channel SNR falls, DIB-PPJSCC is shown to provide a graceful degradation of the reconstruction quality. The overall results showed that the proposed schemes can significantly reduce privacy leakage and improve the reconstruction quality.

\bibliographystyle{IEEEtran}
\bibliography{IEEEabrv}

\begin{thebibliography}{10}
\providecommand{\url}[1]{#1}
\csname url@samestyle\endcsname
\providecommand{\newblock}{\relax}
\providecommand{\bibinfo}[2]{#2}
\providecommand{\BIBentrySTDinterwordspacing}{\spaceskip=0pt\relax}
\providecommand{\BIBentryALTinterwordstretchfactor}{4}
\providecommand{\BIBentryALTinterwordspacing}{\spaceskip=\fontdimen2\font plus
\BIBentryALTinterwordstretchfactor\fontdimen3\font minus
  \fontdimen4\font\relax}
\providecommand{\BIBforeignlanguage}[2]{{%
\expandafter\ifx\csname l@#1\endcsname\relax
\typeout{** WARNING: IEEEtran.bst: No hyphenation pattern has been}%
\typeout{** loaded for the language `#1'. Using the pattern for}%
\typeout{** the default language instead.}%
\else
\language=\csname l@#1\endcsname
\fi
#2}}
\providecommand{\BIBdecl}{\relax}
\BIBdecl

\bibitem{shannon1948mathematical}
C.~E. Shannon, ``A mathematical theory of communication,'' \emph{Bell Syst.
  Techn. J.}, vol.~27, no.~3, pp. 379--423, Jul. 1948.

\bibitem{berlekamp1978inherent}
E.~Berlekamp, R.~McEliece, and H.~Van~Tilborg, ``On the inherent intractability
  of certain coding problems (corresp.),'' \emph{{IEEE} Trans. Inf. Theory},
  vol.~24, no.~3, pp. 384--386, May 1978.

\bibitem{chafii2022ten}
M.~Chafii, L.~Bariah, S.~Muhaidat, and M.~Debbah, ``Ten scientific challenges
  for {6G}: Rethinking the foundations of communications theory,'' Available:
  \url{https://arxiv.org/abs/physics/2207.01843}, 2022.

\bibitem{chen2021distributed}
M.~Chen, D.~G{\"u}nd{\"u}z, K.~Huang, W.~Saad, M.~Bennis, A.~V. Feljan, and
  H.~V. Poor, ``Distributed learning in wireless networks: Recent progress and
  future challenges,'' \emph{{IEEE} J. Sel. Areas Commun.}, vol.~39, no.~12,
  pp. 3579--3605, Oct. 2021.

\bibitem{cheung2000bit}
G.~Cheung and A.~Zakhor, ``Bit allocation for joint source/channel coding of
  scalable video,'' \emph{{IEEE} Trans. Image Process.}, vol.~9, no.~3, pp.
  340--356, Mar. 2000.

\bibitem{heinen2005transactions}
S.~Heinen and P.~Vary, ``Transactions papers source-optimized channel coding
  for digital transmission channels,'' \emph{{IEEE} Trans. Commun.}, vol.~53,
  no.~4, pp. 592--600, Apr. 2005.

\bibitem{gallager1968information}
R.~G. Gallager, \emph{Information {T}heory and {R}eliable
  {C}ommunication}.\hskip 1em plus 0.5em minus 0.4em\relax {S}pringer, 1968,
  vol.~2.

\bibitem{gastpar2003code}
M.~Gastpar, B.~Rimoldi, and M.~Vetterli, ``To code, or not to code: Lossy
  source-channel communication revisited,'' \emph{{IEEE} Trans. Inf. Theory},
  vol.~49, no.~5, pp. 1147--1158, May 2003.

\bibitem{bourtsoulatze2019deep}
E.~Bourtsoulatze, D.~B. Kurka, and D.~G{\"u}nd{\"u}z, ``Deep joint
  source-channel coding for wireless image transmission,'' \emph{{IEEE} Trans.
  Cogn. Commun. Netw.}, vol.~5, no.~3, pp. 567--579, Sept. 2019.

\bibitem{choi2019neural}
K.~Choi, K.~Tatwawadi, A.~Grover, T.~Weissman, and S.~Ermon, ``Neural joint
  source-channel coding,'' in \emph{Proc. Int. Conf. Mach. and Learn.}, Long
  Beach, California, USA, Jun. 2019, pp. 1182--1192.

\bibitem{sun2023adaptive}
L.~Sun, Y.~Yang, M.~Chen, C.~Guo, W.~Saad, and H.~V. Poor, ``Adaptive
  information bottleneck guided joint source and channel coding for image
  transmission,'' \emph{{IEEE} J. Sel. Areas Commun.}, vol.~41, no.~8, pp.
  2628--2644, Aug. 2023.

\bibitem{xu2021wireless}
J.~Xu, B.~Ai, W.~Chen, A.~Yang, P.~Sun, and M.~Rodrigues, ``Wireless image
  transmission using deep source channel coding with attention modules,''
  \emph{{IEEE} Trans. Circuits Syst. Video Technol.}, Apr. 2021.

\bibitem{song2020infomax}
Y.~Song, M.~Xu, L.~Yu, H.~Zhou, S.~Shao, and Y.~Yu, ``Infomax neural joint
  source-channel coding via adversarial bit flip,'' in \emph{Proc. AAAI Conf.
  Artificial Intell.}, New York, USA, Feb. 2020, pp. 5834--5841.

\bibitem{xu2021deep}
J.~Xu, B.~Ai, W.~Chen, N.~Wang, and M.~Rodrigues, ``Deep joint encryption and
  source-channel coding: An image visual protection approach,'' Available:
  \url{https://arxiv.org/abs/2111.03234}, 2021.

\bibitem{tung2022deep}
T.-Y. Tung and D.~Gunduz, ``Deep joint source-channel and encryption coding:
  Secure semantic communication,'' Available:
  \url{https://arxiv.org/abs/2208.09245}, 2022.

\bibitem{massoudi2008overview}
A.~Massoudi, F.~Lefebvre, C.~De~Vleeschouwer, B.~Macq, and J.-J. Quisquater,
  ``Overview on selective encryption of image and video: challenges and
  perspectives,'' \emph{Eurasip J. Inf. Security}, vol. 2008, no.~1, pp. 1--18,
  Nov. 2008.

\bibitem{Marchioro2020Adversarial}
T.~Marchioro, N.~Laurenti, and D.~Gündüz, ``Adversarial networks for secure
  wireless communications,'' in \emph{Proc. Int. Conf. Acoustics Speech Signal
  Process.}, Virtual, May 2020, pp. 8748--8752.

\bibitem{Erdemir2022Privacy}
E.~Erdemir, P.~L. Dragotti, and D.~Gündüz, ``Privacy-aware communication over
  a wiretap channel with generative networks,'' in \emph{Proc. Int. Conf.
  Acoustics Speech Signal Process.}, Shenzhen, China, Oct. 2022, pp.
  2989--2993.

\bibitem{kurka2020deepjscc}
D.~B. Kurka and D.~G{\"u}nd{\"u}z, ``Deep{JSCC}-f: Deep joint source-channel
  coding of images with feedback,'' \emph{{IEEE} J. Sel. Areas Inf. Theory},
  vol.~1, no.~1, pp. 178--193, May 2020.

\bibitem{Roy_2019_CVPR}
P.~C. Roy and V.~N. Boddeti, ``Mitigating information leakage in image
  representations: A maximum entropy approach,'' in \emph{Proc. {IEEE} Conf.
  Comput. Vis. Pattern Recog.}, Long Beach, California, USA, June 2019, pp.
  2586--2594.

\bibitem{BOULEMTAFES202021}
A.~Boulemtafes, A.~Derhab, and Y.~Challal, ``A review of privacy-preserving
  techniques for deep learning,'' \emph{Neur. Comput.}, vol. 384, pp. 21--45,
  Apr. 2020.

\bibitem{bloch2021overview}
M.~Bloch, O.~Günlü, A.~Yener, F.~Oggier, H.~V. Poor, L.~Sankar, and R.~F.
  Schaefer, ``An overview of information-theoretic security and privacy:
  Metrics, limits and applications,'' \emph{{IEEE} J. Sel. Areas Inf. Theory},
  vol.~2, no.~1, pp. 5--22, Mar. 2021.

\bibitem{tishby2000information}
N.~Tishby, F.~C. Pereira, and W.~Bialek, ``The information bottleneck method,''
  Available: \url{https://arxiv.org/abs/physics/0004057}, 2000.

\bibitem{pan2021disentangled}
Z.~Pan, L.~Niu, J.~Zhang, and L.~Zhang, ``Disentangled information
  bottleneck,'' in \emph{Proc. AAAI Conf. Artificial Intell.}, Virtual, Feb.
  2021, pp. 9285--9293.

\bibitem{alemi2016deep}
A.~A. Alemi, I.~Fischer, J.~V. Dillon, and K.~Murphy, ``Deep variational
  information bottleneck,'' Available: \url{https://arxiv.org/abs/1612.00410},
  2016.

\bibitem{pmlr-v80-kim18b}
H.~Kim and A.~Mnih, ``Disentangling by factorising,'' in \emph{Proc. Int. Conf.
  Mach. and Learn.}, Stockholm, Sweden, Jul. 2018, pp. 2649--2658.

\bibitem{NEURIPS2018_1ee3dfcd}
R.~T.~Q. Chen, X.~Li, R.~B. Grosse, and D.~K. Duvenaud, ``Isolating sources of
  disentanglement in variational autoencoders,'' in \emph{Proc. Adv. Neural
  Inform. Process. Syst.}, Montreal, Canada, Dec. 2018, p. 2610–2620.

\bibitem{Chen_2021_ICCV}
Z.~Chen, Y.~Luo, R.~Qiu, S.~Wang, Z.~Huang, J.~Li, and Z.~Zhang, ``Semantics
  disentangling for generalized zero-shot learning,'' in \emph{Proc. Int. Conf.
  Comput. Vis.}, Virtual, Oct. 2021, pp. 8712--8720.

\bibitem{nguyen2010estimating}
X.~Nguyen, M.~J. Wainwright, and M.~I. Jordan, ``Estimating divergence
  functionals and the likelihood ratio by convex risk minimization,''
  \emph{{IEEE} Trans. Inf. Theory}, vol.~56, no.~11, pp. 5847--5861, Nov. 2010.

\bibitem{Hadad2018two}
N.~Hadad, L.~Wolf, and M.~Shahar, ``A two-step disentanglement method,'' in
  \emph{Proc. {IEEE} Conf. Comput. Vis. Pattern Recog.}, Salt Lake City, USA,
  Jun. 2018, pp. 772--780.

\bibitem{Hin2006Fast}
G.~E. Hinton, S.~Osindero, and Y.-W. Teh, ``A fast learning algorithm for deep
  belief nets,'' \emph{Neur. Comput.}, vol.~18, no.~7, pp. 1527--1554, Jul.
  2006.

\bibitem{lecun1998mnist}
E.~Sariyildiz, H.~Yu, and K.~Ohnishi, ``Gradient-based learning applied to
  document recognition,'' \emph{Proceedings of the IEEE}, vol.~86, no.~11, pp.
  2278--2324, Nov. 1998.

\bibitem{Zhang2017Age}
Z.~Zhang, Y.~Song, and H.~Qi, ``Age progression/regression by conditional
  adversarial autoencoder,'' in \emph{Proc. {IEEE} Conf. Comput. Vis. Pattern
  Recog.}, Hawaii, USA, Jul. 2017, pp. 5810--5818.

\bibitem{BPG}
F.~Bellard, ``{BPG} image format,'' Available: \url{https://bellard.org/bpg/},
  2018.

\bibitem{AES}
S.~Heron, ``Advanced encryption standard (aes),'' \emph{Netw. Security}, vol.
  2009, no.~12, pp. 8--12, Dec. 2009.

\bibitem{LDPC}
R.~Gallager, ``Low-density parity-check codes,'' \emph{IRE Trans. Inf. Theory},
  vol.~8, no.~1, pp. 21--28, Jan. 1962.

\bibitem{van2008visualizing}
L.~Van~der Maaten and G.~Hinton, ``Visualizing data using t-{SNE}.'' \emph{J.
  Mach. Learn. Research}, vol.~9, no.~11, Nov. 2008.

\bibitem{hagenauer1996iterative}
J.~Hagenauer, E.~Offer, and L.~Papke, ``Iterative decoding of binary block and
  convolutional codes,'' \emph{{IEEE} Trans. Inf. Theory}, vol.~42, no.~2, pp.
  429--445, Mar. 1996.

\bibitem{vikalo2004iterative}
H.~Vikalo, B.~Hassibi, and T.~Kailath, ``Iterative decoding for {MIMO} channels
  via modified sphere decoding,'' \emph{{IEEE} Trans. Commun.}, vol.~3, no.~6,
  pp. 2299--2311, Spet. 2004.

\end{thebibliography}
\newpage

 




\vfill
\end{document}